\documentclass[prl,
 reprint,
superscriptaddress,
 amsmath,amssymb,
 aps,
]{revtex4-2}
\usepackage{rotating}
\usepackage{multirow}
\usepackage{comment}
\usepackage{graphicx}
\usepackage{dcolumn}
\usepackage{bm}
\usepackage{ulem}
\usepackage[dvipsnames]{xcolor}

\def\half{{\textstyle \frac{1}{2}}}

\def \dLdot {\ensuremath{{\dot{\Delta L}}}}
\def \mubif {\ensuremath{{\mu_\ast}}}
\def \tsnap {0}
\def \tosc {\ensuremath{{t_\mathrm{osc}}}}

\begin{document}

\preprint{APS/123-QED}

\title{Transient amplification of broken symmetry in elastic snap-through}

\author{Qiong Wang}\thanks{These authors contributed equally.}
 \affiliation{Department of Mechanical Science and Engineering, University of Illinois at Urbana-Champaign, Urbana, IL 61801}
 
\author{Andrea Giudici}\thanks{These authors contributed equally.}
\affiliation{Mathematical Institute, University of Oxford, Woodstock Rd, Oxford, OX2 6GG, UK}

\author{Weicheng Huang}
\affiliation{School of Engineering, Newcastle University, Newcastle upon Tyne NE1 7RU, UK}

\author{Yuzhe Wang}
\affiliation{Singapore Institute of Manufacturing Technology, Agency for Science, Technology and Research, Singapore, 138634, Singapore}

\author{Mingchao Liu}
\affiliation{Mathematical Institute, University of Oxford, Woodstock Rd, Oxford, OX2 6GG, UK}
\affiliation{School of Mechanical and Aerospace Engineering, Nanyang Technological University, Singapore 639798, Singapore}
\affiliation{Department of Mechanical Engineering, University of Birmingham, Birmingham, B15 2TT, UK}

\author{Sameh Tawfick}%
\affiliation{Department of Mechanical Science and Engineering, University of Illinois at Urbana-Champaign, Urbana, IL 61801}

\author{Dominic Vella}%
 \email{dominic.vella@maths.ox.ac.uk}
\affiliation{Mathematical Institute, University of Oxford, Woodstock Rd, Oxford, OX2 6GG, UK}

\date{\today}

\begin{abstract}
A snap-through bifurcation occurs when a bistable structure loses one of its stable states and moves rapidly to the remaining state. For example, a buckled arch with symmetrically clamped ends can snap between an inverted and a natural state as the ends are released. A standard linear stability analysis suggests that the arch becomes unstable to asymmetric perturbations. Surprisingly, our experiments show that this is not always the case: symmetric transitions are also observed.  Using experiments, numerics, and a toy model, we show that the symmetry of the transition depends on the rate at which the ends are released with  sufficiently fast loading leading to symmetric snap-through.  Our  toy model reveals that this behavior is caused by a region of the system's state space in which any initial asymmetry  is amplified. The system may not enter  this region when loaded fast (hence remaining symmetric) but will traverse it for some interval of time when loaded slowly, causing a transient amplification of asymmetry. Our toy model suggests that this behavior is not unique to snapping arches, but rather can be observed in dynamical systems where both a saddle-node and a pitchfork bifurcation occur in close proximity.

\end{abstract}

\maketitle

The rapidity of elastic snap-through is striking: just as the snapping shut of the venus flytrap's leaves catches its prey by surprise \cite{Forterre2005}, the jumping disc and variants remain popular toys \cite{Isenberg1987,Ucke2009,Pandey2014}. Nevertheless, detailed analysis has shown that the early dynamics of snap-through  may be slower, and last longer, than naively expected --- critical slowing down is observed as the threshold for snap-through is approached \cite{Gomez2017, Sano2019}, while dynamic changes close to threshold lead to a `delayed' snap-through \cite{Liu2021}.

The surprising features of dynamic snap-through result from the underlying bifurcation structure, which is itself sensitively dependent on the underlying symmetries of the problem, particularly the imposed boundary conditions \cite{Radisson2023,Radisson2023b}. The symmetry (or asymmetry) of the snap-through transition has also been shown to be controlled by geometrical parameters such as the aspect ratio (thickness-to-length) of a buckled beam \cite{Pandey2014,Harvey2015}. However, recent experiments that used the rapidity of snap-through to initiate jumping in an insect-sized robot \cite{Wang2023} also suggested that whether snap-through occurs symmetrically or asymmetrically may vary within the same experimental system (i.e.~with the same geometrical and mechanical properties).  This contradicts standard linear stability analyses, which suggests that snap-through should happen asymmetrically \cite{Gomez2018Thesis}. Understanding the origins (or absence) of asymmetry has important implications for controlling the jumps of these robots (e.g.~whether they jump vertically or sideways), but is more generally important in applications of dynamic snap-through  \cite{Holmes2019}.

\begin{figure}[t!]
\includegraphics[width=\columnwidth]{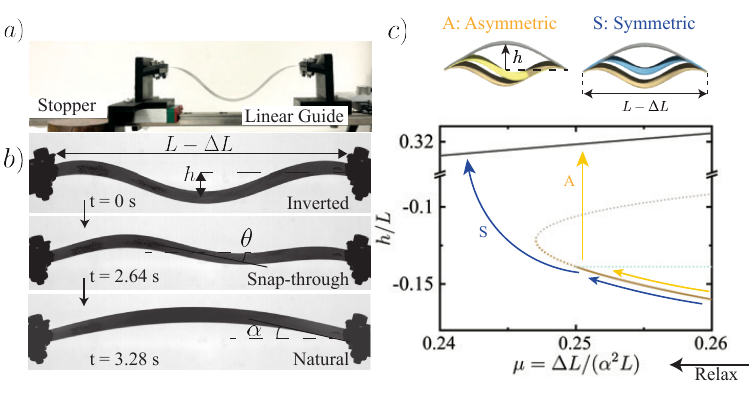}
\caption{
(a) Experimental setup in which the ends of a steel strip (length $L$) are clamped at angle $\alpha$ to the horizontal. The clamp on the right is fixed while that on the left is moved along a linear guide by a motorized stage. (The stopper prevents over-stretching after snap-through.)  (b) Three snapshots of snap-through:  an initial end-end compression, $\Delta L$, is applied allowing buckling in the `inverted' state (top). As $\Delta L$ is slowly decreased, the arch snaps from the inverted state to the natural state (bottom). The angle of inclination at the centre, $\theta(t)$, highlights asymmetry. (c) Bifurcation diagram showing stable equilibria as solid curves: the natural state (black) always exists while the inverted state (brown) vanishes at $\mu=\mubif\approx0.247$ but becomes unstable (dashed curve) for $\mubif\leq\mu<\mu_A=1/4$ because of the presence of an unstable asymmetric mode (cyan dashed line). This complexity gives two routes to snap-through from the inverted to natural states: a symmetric path (blue arrow) and an asymmetric path (gold arrow).}
\label{fig:setup}
\end{figure}

In this Letter, we study the snap-through that powered the jumping of insect-scale robots using a model system: a strip of spring steel (Wear-Resistant 1095 Spring Steel strip, modulus $E\approx 200\mathrm{~GPa}$, thickness $b=0.05\mathrm{~mm}$, width $w=25\mathrm{~mm}$) of length $L=108\mathrm{~mm}$ was formed into an arch by clamping its ends symmetrically at an angle $\alpha=\pi/12\mathrm{~rad}$ to the horizontal and imposing an end-shortening $\Delta L$ on one side (see fig.~\ref{fig:setup}a). Given a clamping angle $\alpha>0$ and sufficiently large values of $\Delta L/L$  the arch can exist in one of two (symmetric) states: the `natural' state (resembling a `$\Lambda$') and an  `inverted' state (resembling an `M')  --- it is bistable \cite{Gomez2018Thesis}. However, as $\Delta L$ decreases, the inverted state ceases to exist and the system is forced to snap to the natural state.   We vary $\Delta L$ experimentally by fixing one clamp while the other is attached to a motorized linear stage. The speed $|\dLdot|$ can be controlled in the range $1 \mathrm{~mm\,s^{-1}} \leq |\dLdot| \leq 30 \mathrm{~cm\,s^{-1}}$.

\begin{figure*}
\includegraphics[width=\textwidth]{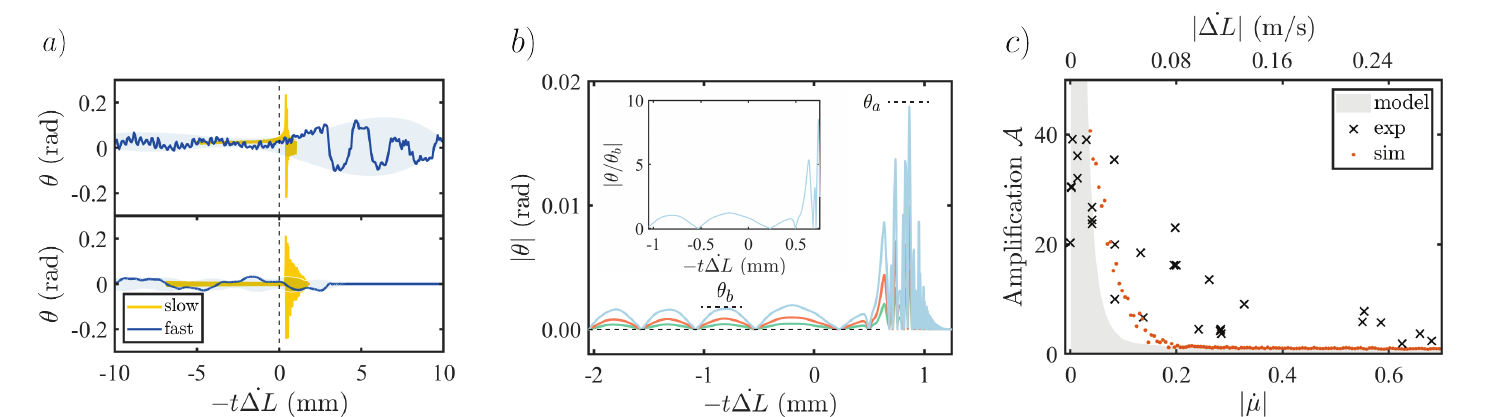}
\caption{Transient asymmetry in the snap-through of a symmetric arch. (a) Both the experimental (top) and simulation (bottom) results show that the central angle $\theta$ grows close to the snap-through transition at $t=0$ (vertical dashed line), but only when the loading rate $|\dot{\Delta L}|$ is small enough. (In the slow case, $|\dLdot|=1.8 \mathrm{~mm/s}$; in the fast case, $|\dLdot|=246.6 \mathrm{~mm/s}$.) The shaded region in each data set shows the envelope of oscillations before and after snap-through. (b) Numerical simulations highlight the role of loading protocol: when both ends are moved with the same acceleration (dashed black line) the system remains perfectly symmetrical. With only one end moved (as in experiments), asymmetric oscillations are excited at low (green), medium (red), and high (blue) accelerations, and thus symmetry is broken. In each case, while the raw values of $\theta$ vary with the driving acceleration, plotting the amplification,  collapses the curves (see inset). (c) The maximum amplification, defined in \eqref{eqn:AmpDefn}, shows a sensitive dependence on the loading rate $|\dot{\Delta L}|$ and its dimensionless equivalent, $\dot{\mu}$, in both experiments and simulations.}
\label{fig:RawData} 
\end{figure*}

`Snap-through' is captured via a high-speed movie (taken at 2245 fps). The images in fig.~\ref{fig:setup}b show that, at least in some circumstances, the arch may snap-through asymmetrically:  the inclination angle at the centre, $\theta(t)$,  differs visibly from zero. Quantitative plots of  $\theta(t)$ are shown in fig.~\ref{fig:RawData}a and highlight that when $\Delta L$ is changed slowly,  an initially small $\theta$ is significantly amplified close to the snap-through transition. Surprisingly, however, this is different when the motion is driven 'quickly': for larger $|\dLdot|$, $\theta$ does not increase significantly.

As a first indication of the root of this rich phenomenology, a bifurcation analysis \footnote{\label{SI} See accompanying Supplementary Information [url] for further details of the experimental, numerical and analytical methods, including Refs \cite{Lowess,Audoly2010}.} reveals several important insights. Firstly, for sufficiently small values of $\alpha>0$, the behavior of the system is controlled by the parameter
\begin{equation}
    \mu=\frac{1}{\alpha^2}\frac{\Delta L}{L},
    \label{eqn:muDef}
\end{equation} which compares the typical angle induced by compression, $(\Delta L/L)^{1/2}$, to the clamping angle, $\alpha$. A static analysis reveals that for $\mu<\mubif\approx0.247...$ the natural state remains as an equilibrium but the inverted state vanishes via a saddle-node bifurcation at $\mu=\mubif$. A standard linear stability analysis \cite{Gomez2018Thesis,Note1} also shows that at $\mu=\mu_A=1/4>\mubif$ the inverted state becomes unstable and,  further, that (linear) instability progresses via a growing asymmetric mode. This behavior, summarized in fig.~\ref{fig:setup}c, leads us to hypothesize that paths in phase space may be symmetric or asymmetric depending on the loading rate (shown schematically in fig.~\ref{fig:setup}c).

To test the robustness of the speed-dependent symmetry/asymmetry observed experimentally, we use numerical experiments with the planar discrete elastic rod method \cite{bergou2008discrete,bergou2010discrete,huang2019newmark}. The physical parameters  in simulations are selected to  correspond to physical experiments; typical results are shown in fig.~\ref{fig:RawData}b and confirm that significant asymmetry can be observed with perfectly symmetrical boundary conditions, but only with sufficiently small $|\dLdot|$. Moreover, numerics show that small oscillations in $\theta$ are observed \emph{before} snap-through; we call these precursor oscillations. 

The numerical experiments show that small precursor oscillations  are vital to asymmetry: when moving both ends perfectly symmetrically, no precursor oscillations are observed and the arch remains symmetric throughout snap-through  (horizontal line in fig.~\ref{fig:RawData}c).  However, moving just one end introduces small asymmetric precursor oscillations in the arch shape (despite the symmetric boundary conditions); these are then amplified during snap-through. Different accelerations excite precursor oscillations of different magnitude (and hence different raw asymmetries during snap-through), as might be expected. However, rescaling by the size of the oscillations before snap-through, $\theta_b=\max_{t<\tsnap}|\theta(t)|$ where $t=\tsnap$ is defined such that $\mu(\tsnap)=\mubif$, allows us to collapse these data (see inset of fig.~\ref{fig:RawData}c). This motivates measuring the amplification of asymmetry, defined as
    \begin{equation}
        {\cal A}:=\frac{\max_{t>\tsnap}|\theta(t)|}{\max_{t<\tsnap}|\theta(t)|}.
        \label{eqn:AmpDefn}
    \end{equation} (For the experiments, the oscillation amplitude  is determined as the standard deviation of $\theta$ for $t<0$ \cite{Note1}.)
Both numerical and physical experiments show a distinct effect of the stretching rate $|\dLdot|$ on ${\cal A}$: \emph{the amount of amplification decreases with increasing end-stretching rate} $|\dLdot|$ (fig.~\ref{fig:RawData}d). Both also pose a series of questions. Firstly, why and how does the end-stretching rate determine the symmetry of the arch as it snaps through?  Secondly, why is this transition observed as an amplification of an existing asymmetry?

\begin{figure}[t]
\includegraphics[width= \columnwidth]{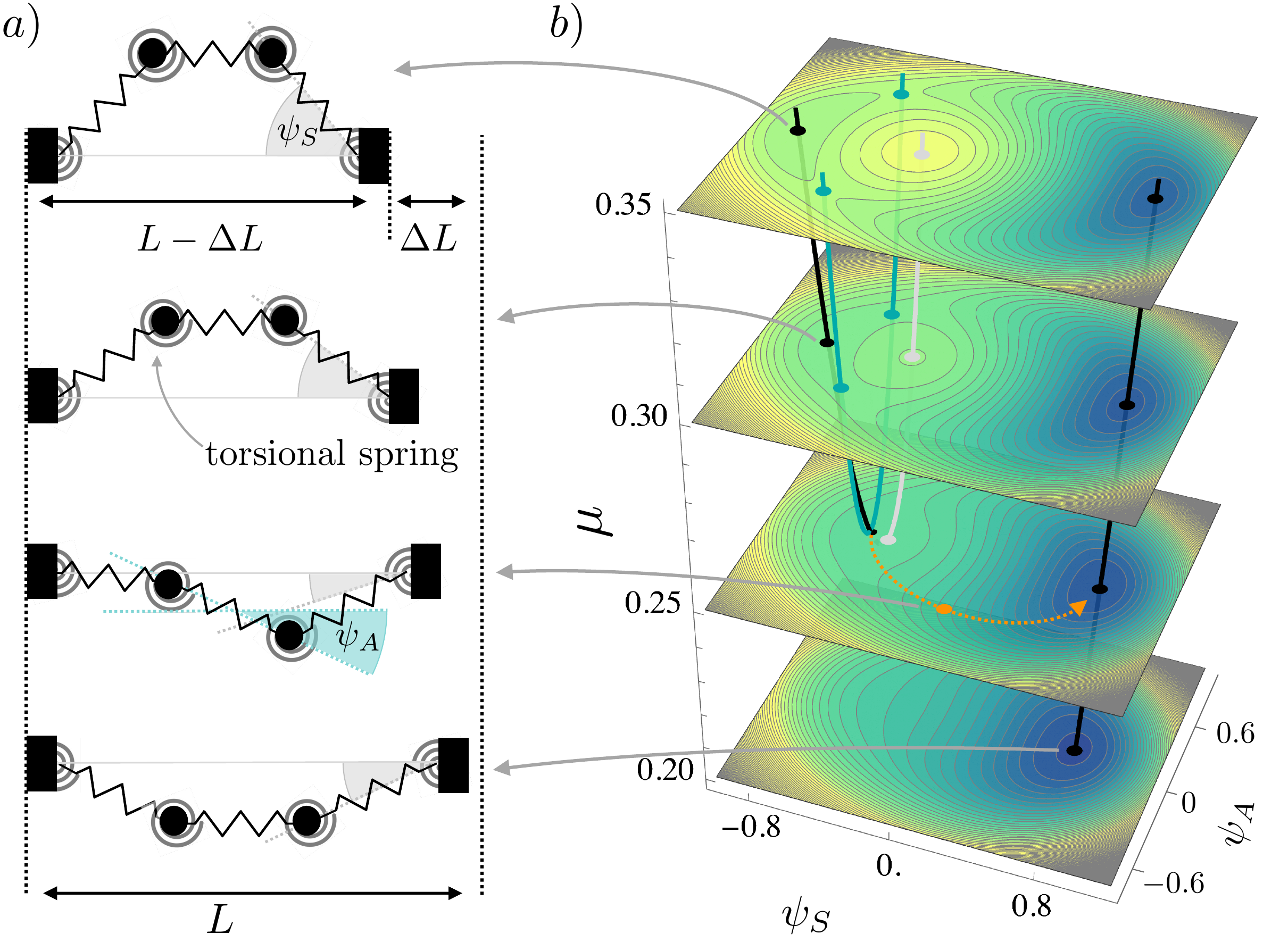}
\caption{\label{fig:energy} (a) A sketch of the double-mass von Mises truss with torsional spring elements in various states, corresponding to the inverted arch  (top) and the natural arch (bottom). (b) The evolution of the energy landscape as a function of the end-shortening parameter $\mu=\Delta L/(L\alpha^2)$.}
\end{figure}

A  simple model to study snap-through instabilities is the von Mises truss \cite{Mises1923,Timoshenko1961}. However, the classic von Mises truss has only one degree of freedom and does not admit asymmetric modes. To capture asymmetry, we introduce the double-mass von Mises truss in which two masses are connected (via linear springs) to two rigid end-points (see fig.~\ref{fig:energy}a). With imposed end-shortening, this system buckles either up or down to reduce the compression of the springs. To bias the system and ensure that one of these buckled states is preferred (the `natural' state), we include torsional springs between the clamps and the side springs whose relaxed position is at an angle $\alpha$ with respect to the horizontal. To stabilize the inverted state, we also include torsional springs connecting the central spring to its neighbors,  penalizing deviations from flat --- this introduces an effective bending stiffness.

A detailed analysis  \cite{Note1}  shows that for small angles $\alpha\ll1$ the state of the system can be described using two angles alone: the symmetric and asymmetric parts of the angles, denoted $\psi_S$ and $\psi_A$, respectively. These angles are illustrated schematically in fig.~\ref{fig:energy}a (see \cite{Note1} for formal definitions); loosely speaking, $\psi_S$ corresponds to the height, $h$, of the corresponding arch while $\psi_A$ is a measure of the asymmetry. For simplicity, we argue via symmetry in the main text, giving details of a formal analysis elsewhere \cite{Note1}. Crucially, the angles $\psi_A, \psi_S \sim \alpha\ll1$ while the relative end-shortening $\Delta L/L \sim \alpha^2$, again motivating $\mu=\Delta L/(\alpha^2L)$. Moreover, when $\alpha=0$ we expect the energy to be even in $\psi_S$. However, the bias introduced by the clamps is linear in $\psi_S$ (and independent of  $\Delta L/L$); as such,  there is an additional linear term in $\psi_S$. Similarly, the energy must be even in $\psi_A$, since the system is left-right symmetric.  Thus, to fourth order in small quantities (i.e.~at $O(\alpha^4)$), the energy takes the form:
\begin{align}
    U = & \; U_0 - a_1 \psi_S + \frac{1}{2} (a_2 - a_3 \mu) \psi_S^2 + \frac{1}{4} a_4 \psi_S^4\nonumber\\
    &+ 
    \frac{1}{2}f(\psi_S; \mu)\psi_A^2 + \frac{1}{4} b_4 \psi_A^4\
    \label{eqn:EnergyLandau}
\end{align}
where
\begin{equation}
f(\psi_S; \mu) = b_1 - b_2 \mu + b_3 \psi_S^2
\label{eqn:DefnF}
\end{equation} and $a_i,b_i>0$ are constants that depend on the initial geometry and mechanical properties of the system \cite{Note1}. The energy in \eqref{eqn:EnergyLandau} can be derived from first principles by summing the energies stored in the linear and torsional springs as functions of the masses' positions subject to a constraint on the total end-displacement. Taking the leading order terms in small clamp angle, $\alpha$, yields \eqref{eqn:EnergyLandau} \cite{Note1}.
Equation \eqref{eqn:EnergyLandau} shows that the coefficients of the two quadratic terms are not constant: the coefficient of $\psi_S^2$ varies with  $\mu$ (as must be the case for a bifurcation to occur) while the coefficient of $\psi_A^2$ depends on $\psi_S$.

 The energy $U$ is plotted as a function of  $\mu$, $\psi_S$ and $\psi_A$ in fig.~\ref{fig:energy}b. When $\mu$ is large, the system has $2$ stable equilibrium states (black points), corresponding to the natural (global minimum) and inverted (local minimum) symmetric states. There are three more equilibria corresponding to a symmetric local maximum (gray) and two asymmetric saddles (cyan). As  $\mu$ decreases, the two asymmetric saddles annihilate when $f(\psi_S;\mu)=0$ at a critical $\mu=\mu_A^{\mathrm{vM}}$. For $\mu<\mu_A^{\mathrm{vM}}$, the inverted equilibrium goes from being a minimum to being a saddle --- it becomes susceptible to asymmetric perturbations. Moreover, when $\mu=\mubif^{\mathrm{vM}}=a_2/a_3+3\left(2 a_1^2 a_4\right)^{1/3}/2a_3$, the inverted equilibrium is annihilated simultaneously with the maximum at $  \psi_S^*=\left(a_1/2 a_4\right)^{1/3}$ in a saddle-node bifurcation; only the natural symmetric equilibrium exists for $\mu<\mu^*$.  Therefore, as  $\mu$ decreases, the inverted equilibrium first becomes unstable to asymmetric perturbations before ceasing to exist altogether. This model replicates the bifurcation structure of the  continuous arch (see fig.~\ref{fig:setup}c) but does so with only two variables,  $(\psi_S,\psi_A)$, rather than the continuous arch shape.  Furthermore, by appropriate choice of two parameters (corresponding to the stiffness ratio of the  torsional and linear springs and the length ratio of the central and edge springs), the two bifurcation points can be tuned to match those of the continuous arch --- we  ensure that $\mubif^{\mathrm{vM}}=\mubif\approx0.247$ and $\mu_A^{\mathrm{vM}}=\mu_A=1/4$ and omit the superscript $^{\mathrm{vM}}$ henceforth.

To understand the dynamic behavior, and further quantify the energetic picture above, we use a variational approach to derive equations for the evolution of $\psi_S$ and $\psi_A$.
Ignoring higher order terms and rescaling time by the period of oscillations, $\tosc \left[m/(\alpha^2k)\right]^{1/2}$, where $\tosc=9.477$  \cite{Note1}, the kinetic energy of the system can be written as $T= \dot{\psi_S}^2+\tfrac{1}{4}\dot{\psi_A}^2$, with differing prefactors between the translational energy of the centre of mass and the rotational energy  \cite{Note1}. A standard application of the Euler--Lagrange equations then gives that, for $\psi_A\ll1$
\begin{align}
\label{eqn:dyn1}
    \ddot{\psi}_S/{\tosc}^2&=\half a_1 - \half(a_2-a_3 \mu)\psi_S- \half a_4 \psi_S^3\\
        \label{eqn:dyn2}
    \ddot{\psi}_A/{\tosc}^2&=-2f(\psi_S;\mu) \psi_A
\end{align}
 
 The system of equations \eqref{eqn:dyn1}--\eqref{eqn:dyn2} yield considerable insight. Firstly,  \eqref{eqn:dyn2}  illustrates a one-way coupling from $\psi_S$ to $\psi_A$: the evolution of $\psi_A$ depends on $\psi_S$ but $\psi_S$ evolves independently of $\psi_A$. More importantly, the equilibrium value of the function $f(\psi_S;\mu)$, in \eqref{eqn:dyn2}, changes sign when $\mu=\mu_A=1/4$. For $\mu>\mu_A$, any small $\psi_A$ exhibits simple harmonic motion; for $\mu<\mu_A$ the oscillatory behavior of $\psi_A$ changes to exponential growth and the symmetric equilibrium, $\psi_A=0$, is unstable. Moreover, \eqref{eqn:dyn1} loses a pair of equilibrium solutions (a saddle-node bifurcation \cite{Strogatz2015}) at $\mu=\mubif\approx0.247$. 
 This reproduces the bifurcation picture in fig.~\ref{fig:setup}c.   
 
\begin{figure}[t!]
\includegraphics[width=\columnwidth]{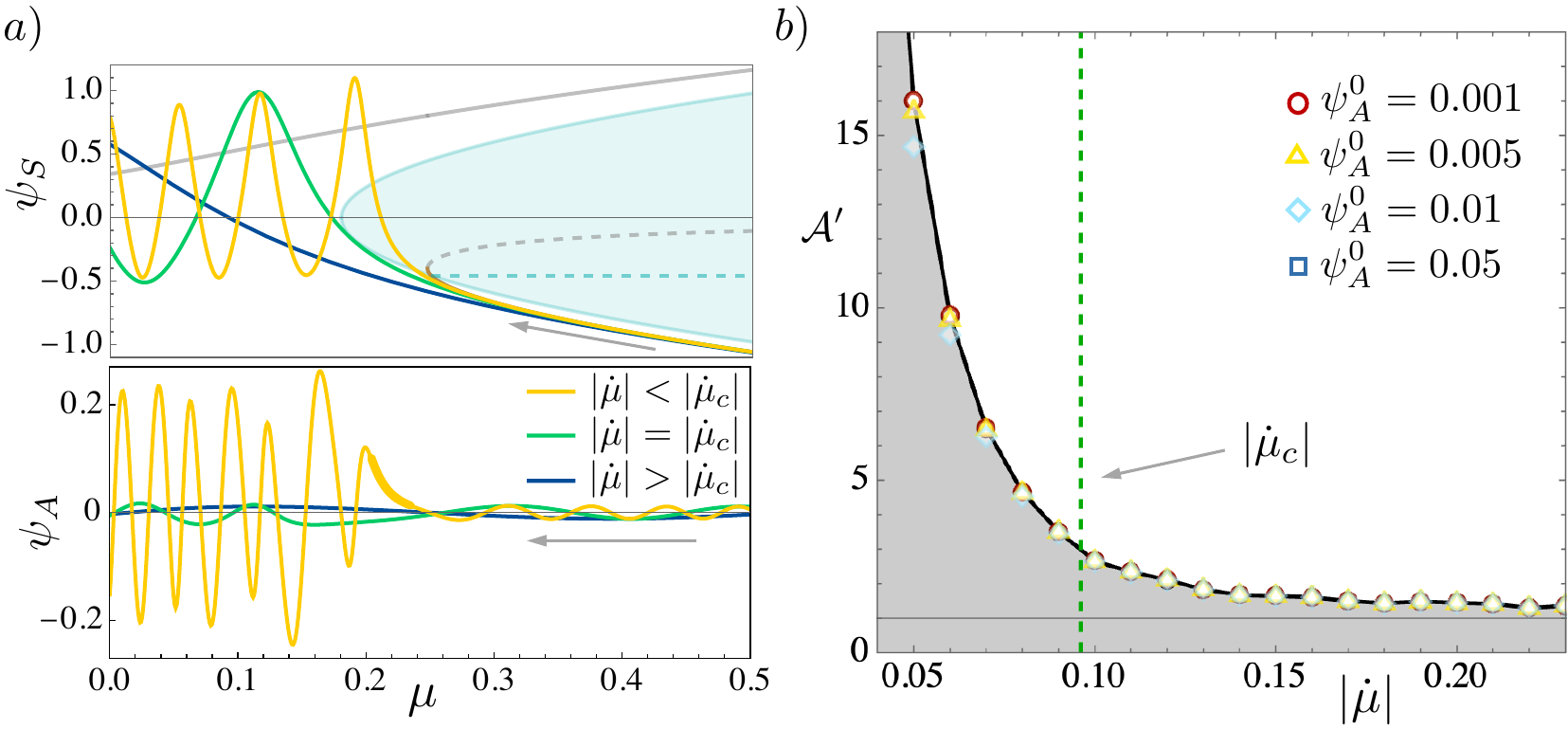}
\caption{(a) Trajectories of $\psi_S$ and $\psi_A$ for the double-mass von Mises truss model for different values of the loading rate $|\dot{\mu}|$ with an initial perturbation $\psi_A^0=0.01$. (Note that time goes from right to left here, since $\dot{\mu}<0$.) The blue shaded region highlights the amplification region in which $f(\psi_S;\mu)<0$: while a trajectory crosses this region, any asymmetry is amplified --- see \eqref{eqn:dyn2} --- highlighted by thicker lines in the trajectory. This is observed only for $|\dot{\mu}|<|\dot{\mu}_c|$.
(b) Numerical results for the fully nonlinear double-mass von Mises truss model show that results with different initial asymmetry can be collapsed by considering the maximum amplification ${\cal A}' := \max{|\psi_A|}/|\psi_A^0|$ and that amplification only occurs for sufficiently slow loading, $|\dot{\mu}|<|\dot{\mu}_c|\approx0.0961$  (vertical dashed line).}
\label{fig:dynamics} 
\end{figure}

 The change in behavior as $f$ changes sign in \eqref{eqn:dyn2} explains our observation that asymmetry in the initial conditions is amplified.
 Since amplification only occurs within the region of parameter space for which $f(\psi_S;\mu)<0$, we refer to this region as the \emph{amplification region} --- shown as the shaded region in the top panel of fig.~\ref{fig:dynamics}a. Whether a trajectory enters the amplification region depends on the evolution of $\psi_S$, which is governed by \eqref{eqn:dyn1} and depends not only on $t$ and the instantaneous value of $\mu$, but also on the rate at which $\mu$ changes, $|\dot{\mu}|$. We write $$\mu=\mubif-|\dot{\mu}| t$$ where $t$ is the rescaled time and highlight that $\psi_S=\psi_S(t;|\dot{\mu}|)$. For small $|\dot{\mu}|$, the trajectory passes very close to the equilibrium solution, effectively evolving quasi-statically with only a rapid change in $\psi_S$ as oscillations in $\psi_A$ give way to exponential growth. For larger values of $|\dot{\mu}|$, however, the trajectory passes further off the equilibrium path and the fast change in $\psi_S$ associated with snap-through occurs with $\mu<\mubif\approx0.247$  --- a phenomenon known as `delayed bifurcation' \cite{Su2001,Liu2021}.

 Crucially, the dependence of the trajectory on  $|\dot{\mu}|$ affects whether the trajectory enters the amplification region at all: for small $|\dot{\mu}|$, trajectories \emph{do} enter the amplification region (thick portion of the yellow/light trajectory in fig.~\ref{fig:dynamics}a) and a small initial value, $\psi_A(0)=\psi_A^0$, is amplified as the trajectory passes through the amplification region. For large $|\dot{\mu}|$, trajectories do \emph{not} pass through the amplification region and $\psi_A$ remains small throughout (dark blue trajectories in fig.~\ref{fig:dynamics}a). This explains why there is a sharp change in behavior: significant asymmetry is observed only for sufficiently small $|\dot{\mu}|$.

 To determine the critical value of $|\dot{\mu}|$, i.e.~what we mean by sufficiently slow/fast, we must determine the value of $|\dot{\mu}|$ for which the trajectory just grazes the amplification region.  The key result of this analysis \cite{Note1} is that the critical value $|\dot{\mu}_c|\approx0.0961$. This is compared with numerical results for the (full) system in fig.~\ref{fig:dynamics}b --- points show the numerically-observed amplification,  while the prediction $|\dot{\mu}|=|\dot{\mu}_c|$ is shown by the vertical dashed line. As expected, the results show very little amplification for $|\dot{\mu}|>|\dot{\mu}_c|$. Moreover, fig.~\ref{fig:dynamics}b shows that for $|\dot{\mu}|<|\dot{\mu}_c|$ there is significant amplification of the underlying asymmetry, and hence an appreciable asymmetry develops. This can be understood heuristically as resulting from the transient nature of the amplification: asymmetry is only amplified while the trajectory passes through the amplification region. Trajectories with small $|\dot{\mu}|$ spend longer  passing through this region, and hence experience greater amplification.  One subtlety is that the phase of oscillation as  snap-through occurs plays a role in determining the precise degree of amplification that is observed \cite{Note1}. Hence, the shaded region in figs \ref{fig:RawData}c and \ref{fig:dynamics}b show the range of amplifications possible as the phase varies.

    In this Letter, we have presented  experimental and numerical evidence that elastic snap-through may take place symmetrically or asymmetrically depending on the rate at which the system is driven through the snap-through transition.  The double-mass von Mises truss model we have presented is thus able to capture the key physics that governs the behavior of a snapping arch observed in experiments and numerics. In particular, it shows that the asymmetry observed is caused by the transient amplification of a small initial asymmetric perturbation. Further, the model shows that this amplification occurs only when the dynamic loading is slow enough that the trajectory of the system passes through an \textit{amplification zone}, thereby explaining the sharp transition in behavior that is observed at a critical loading rate $|\dot{\mu}_c|$.     Qualitatively the results of this toy model agree with the experimental and numerical results (compare fig.~\ref{fig:RawData}d and fig.~\ref{fig:dynamics}b).

The behavior we have reported in this Letter is general: it occurs whenever a system has both a saddle-node and a pitchfork bifurcation, since then its energy has the structure of \eqref{eqn:EnergyLandau}, specifically a term quadratic in $\psi_A$ whose prefactor depends on $\psi_S$. The amplification of the variable associated with the pitchfork bifurcation ($\psi_A$ in our example), depends entirely on $f$ and can be observed provided the analog of $f<0$ at the saddle-node point $\mu_*$ for some values of $\psi_S$. The behavior of such systems close to the bifurcation point cannot be understood from a simple linear stability analysis --- \emph{how} the instability develops depends on fundamentally nonlinear phenomena.     The generic nature of the transient amplification of asymmetry suggests the use of snap-through stimulation \emph{rate} for control of elastic instabilities. For example, in the snap-induced jumping of robots \cite{Wang2023}, the ability to controllably switch snap-through between symmetric and asymmetric modes may lead to the ability to switch between vertical and forwards jumps.  

Our numerics and toy model focus  on the situation in which the boundary conditions are perfectly symmetric. However, in any experiment or application, small defects or imperfections are unavoidable. For small defects, the picture we have presented here persists, while for larger defects the rate of loading plays a slightly different role \cite{Giudici2024}.

\begin{acknowledgments} 
This work was partially supported by the UK Engineering and Physical Sciences Research Council via grant EP/W016249/1 (D.V.). Q.W., Y.W., and S.T.~acknowledge support from Defense Advanced Research Projects Agency DARPA SHRIMP HR001119C0042. M.L.~ acknowledges support via start-up funding from the University of Birmingham.  We are grateful to Michael Gomez for discussions.
\end{acknowledgments}


\providecommand{\noopsort}[1]{}\providecommand{\singleletter}[1]{#1}%

\end{document}






\title{Supplemental Information ``Transient amplification of broken \\ symmetry in elastic snap-through"}

\author{Qiong Wang}\thanks{These authors contributed equally}
 \affiliation{Department of Mechanical Science and Engineering, University of Illinois at Urbana-Champaign, Urbana, IL 61801}
 
\author{Andrea Giudici}\thanks{These authors contributed equally}
\affiliation{Mathematical Institute, University of Oxford, Woodstock Rd, Oxford, OX2 6GG, UK}

\author{Weicheng Huang}
\affiliation{School of Engineering, Newcastle University, Newcastle upon Tyne NE1 7RU, UK}

\author{Yuzhe Wang}
\affiliation{Singapore Institute of Manufacturing Technology, Agency for Science, Technology and Research, Singapore, 138634, Singapore}

\author{Mingchao Liu}
\affiliation{Mathematical Institute, University of Oxford, Woodstock Rd, Oxford, OX2 6GG, UK}
\affiliation{School of Mechanical and Aerospace Engineering, Nanyang Technological University, Singapore 639798, Singapore}
\affiliation{Department of Mechanical Engineering, University of Birmingham, Birmingham, B15 2TT, UK}

\author{Sameh Tawfick}%
\affiliation{Department of Mechanical Science and Engineering, University of Illinois at Urbana-Champaign, Urbana, IL 61801}

\author{Dominic Vella}%
 \email{dominic.vella@maths.ox.ac.uk}
\affiliation{Mathematical Institute, University of Oxford, Woodstock Rd, Oxford, OX2 6GG, UK}

\maketitle

In this Supplementary Information, we present further details of the experimental setup and further experimental results (\S\ref{sec:SI:ExpAll}), the analysis that leads to the bifurcation diagram in fig.~1c of the main paper (\S\ref{sec:SI:Bifurcation}) and the numerical simulations (\S\ref{sec:SI:Nums}). The bulk of the SI, however, focuses on the development of the toy model (\S\ref{sec:SI:Toy}), including giving further details of the various constants and the simplifications that arise when the angle $\alpha\ll1$.

\section{Experiments\label{sec:SI:ExpAll}}
\subsection{Experimental setup and analysis\label{sec:SI:Expt}}

During the experiment, a strip of spring steel (from \textit{McMASTER-CARR}, Wear-Resistant 1095 Spring Steel strip, modulus $E\approx 200\mathrm{~GPa}$, thickness $b=0.05\mathrm{~mm}$, width $w=25\mathrm{~mm}$)  of length $L=108\mathrm{~mm}$ was formed into an arch by clamping its ends symmetrically at an angle $\alpha=\pi/12\mathrm{~rad}$ to the horizontal using two holders.

\begin{figure}[t]
\includegraphics[width=\columnwidth]{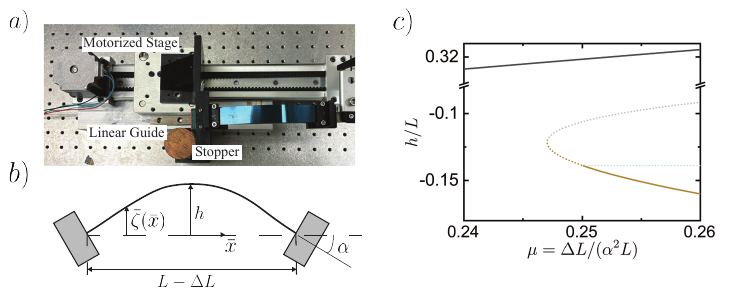}
\caption{(a) Top view of the experimental setup showing the steel strip (lower right). (b)The geometry of a strip with clamped-clamped boundary conditions. (c) Bifurcation diagram showing the height of the arch center, $h=\bar{\zeta}(0)$, as a function of $\mu$ in equilibrium. Symmetric modes with the lowest three compressive forces are plotted as the black (natural mode), brown (inverted mode) and grey curves. The asymmetric mode with the lowest compressive force is shown as dashed cyan line. Stable solutions are shown as full curves, unstable as dashed.  }
\label{fig:ExpBif}
\end{figure}

The loading setup comprises a linear guide, a linear motorized stage, and a stopper, as shown in fig.~\ref{fig:ExpBif}a. This ensures that the holders can be moved relative to one another: the holder on the right side is fixed in the laboratory frame, while the holder on the left side is connected to the motorized linear stage via a piece of double-sided tape. As the stage moves, the holder moves with it until it hits the stopper, at which point it detaches from the moving stage. (The stopper is placed just beyond the position at which snap-through is expected. This position is determined by moving the left holder quasistatically until  snap-through is observed.) Videos of the snap-through process were taken with a high-speed camera (Fastec Hispec 5). The images in the video contained $1184 \mathrm{~pixels} \times 424\mathrm{~pixels}$;  the resolution is $96 \mathrm{~pix/in} \times96 \mathrm{~pix/in}$.

The driving frequency and acceleration of the motor that drives the stage are pre-programmed; the frequency ranged from $10 \mathrm{~Hz}$ to $10^5\mathrm{~Hz}$ while the acceleration time was fixed at the minimum setting ($0.01 \mathrm{~s}$). Within one test, the motor first accelerates, taking the left holder with it and decreasing the end shortening $\Delta L$. As $\Delta L$ further decreases and passes the critical value where the snap-through happens, the holder hits the stopper and detaches from the motor while the motor continues to move before decelerating. 
All videos are post-processed in Matlab. The grayscale images are binarized and the strip is identified using the \textit{bwboundaries} function. After extracting the shape of the strip from the videos, the shape data of the strip is smoothed by \textit{Lowess Smoothing} \cite{Lowess}. 

\subsection{Experimental Results\label{sec:SI:Results}}

\begin{figure}[t]
\includegraphics[width=\columnwidth]{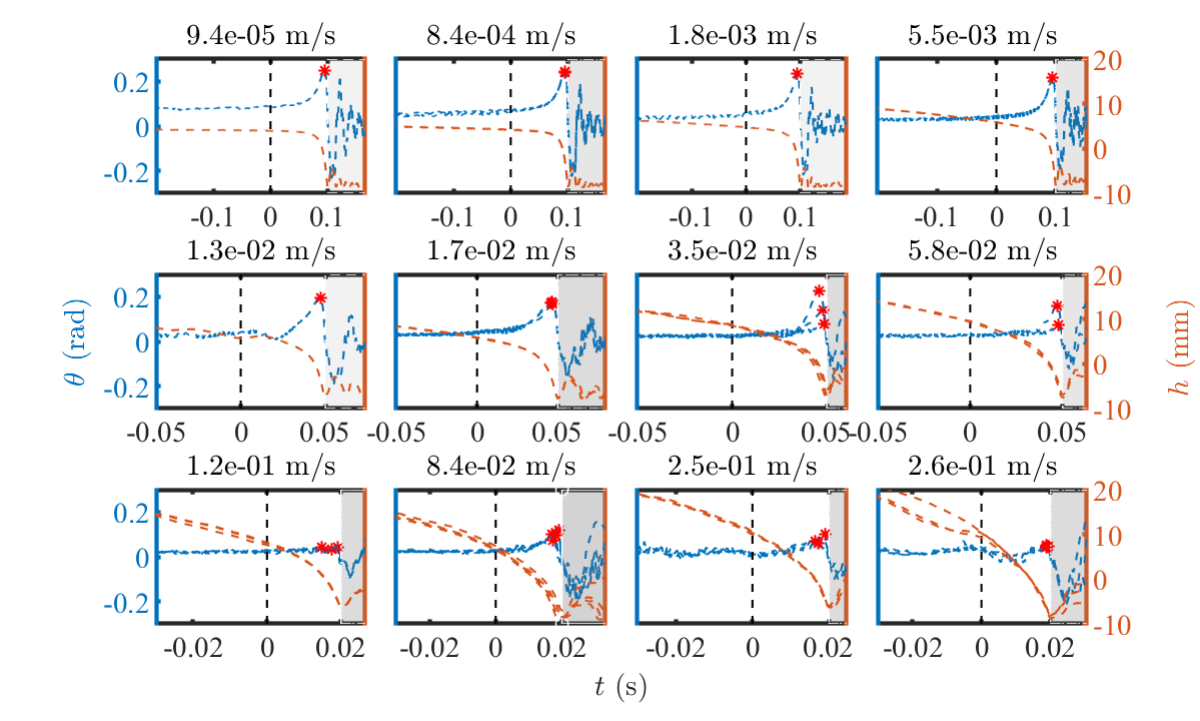}
\caption{The angle of inclination at the centre, $\theta(t)$ and the midpoint deflection $h(t)$ versus time $t$ for tests with 12 different end-point speeds. The black dashed line indicates $t=0$ while the red star highlights the maximum value of $\theta$ measured during snap-through. The greyed area indicates the post-snapping region, $t>t_0$, which is not included in the analysis.}
\label{fig:ExpResults}
\end{figure}

We conducted 30 experiments with different driving speeds, $\dLdot$, using the experimental setup described in \S\ref{sec:SI:Expt}. The speed of the moving clamp is actively controlled by the motor frequencies and directly measured by calibrated high-speed videos. Typical evolution of the mid-point angle $\theta$ and mid-point height $h$ are shown in fig.~\ref{fig:ExpResults} for a range of driving frequencies (and hence measured values of $\dLdot$). In fig.~\ref{fig:ExpResults}, the data is cropped following the snap-through point to better demonstrate the transition. For all experiments, the measured speed of the test and the measured amplification of asymmetry are listed in table \ref{table:TestResults}.

Within each experiment, the timestamp is synced by aligning with the moment when the mid-point height $h$ reaches its first local minimum after snap-through is initiated, denoted as $t_0$. The data for $t>t_0$ is described as post-snap-through behavior and is not considered in this analysis as shown in the greyed area in fig.~\ref{fig:ExpResults}). To quantify the amplification of the oscillations during the snap-through process in the experiments, we take the following measures. The amplitude of the oscillation before snap-through is estimated by $\sqrt{2}\,\mathrm{std}_{t_{-1}<t<0}|\theta(t)|$ where $t=0$ is the time of the point at which snap-through begins and $t_{-1}$ is the point where the evaluation started (e.g.~$t_{-1}=-1\mathrm{~s}$ for the test with speed $9.4\times10^{-5} \mathrm{~m\,s^{-1}}$). The values of $t_{-1}$ and $t_0$ are  selected manually for each test and their variation do not impact the measured amplification. The amplitude of the oscillation during snap-through is measured by the difference between the average value of $\theta$ before snap-through and the maximum value of $\theta$ during snap-through (denoted by red star markers in fig.~\ref{fig:ExpResults}). Then the amplification ${\cal A}$ is given by

\begin{equation}
        {\cal A}:=\frac{\max_{0<t<t_0}|\theta(t)|-\mathrm{mean}_{t_1<t<0}|\theta(t)|}{\sqrt{2}\,\mathrm{std}_{t_1<t<0
        }|\theta(t)|}.
        \label{eqn:AmpDefn}
    \end{equation}
This expression for the amplification in experiments captures the concept of amplification as described in Equation (2) of the main text, but modified to be robust to noise and small imperfections.

\begin{table}[]
\begin{tabular}{||ccc|ccc|ccc||}
\hline
\#  & $|\dLdot|$ (m/s) & {\cal A} & \#  & $|\dLdot|$ (m/s) & {\cal A} & \#  &  $|\dLdot|$ (m/s) & {\cal A} \\ \hline \hline 
1                           & 0.000094                         & 20.31                             & 11                          & 0.035                           & 35.40                             & 21                          & 0.12                            & 4.18                              \\ \hline
2                           & 0.00084                         & 30.32                             & 12                          & 0.035                           & 10.02                             & 22                          & 0.12                            & 4.53                              \\ \hline
3                           & 0.00084                         & 30.53                             & 13                          & 0.035                           & 19.95                             & 23                          & 0.12                            & 3.74                              \\ \hline
4                           & 0.0018                          & 39.22                             & 14                          & 0.056                           & 18.44                             & 24                          & 0.14                            & 9.11                              \\ \hline
5                           & 0.0055                          & 36.14                             & 15                          & 0.058                           & 6.67                              & 25                          & 0.23                            & 5.87                              \\ \hline
6                           & 0.0055                          & 32.09                             & 16                          & 0.082                           & 16.19                             & 26                          & 0.23                            & 7.76                              \\ \hline
7                           & 0.013                           & 39.11                             & 17                          & 0.083                           & 23.07                             & 27                          & 0.25                            & 5.70                              \\ \hline
8                           & 0.017                           & 24.40                             & 18                          & 0.084                           & 16.22                             & 28                          & 0.26                            & 1.85                              \\ \hline
9                           & 0.017                           & 26.76                             & 19                          & 0.10                            & 4.52                              & 29                          & 0.28                            & 3.70                              \\ \hline
10                          & 0.017                           & 23.71                             & 20                          & 0.11                            & 13.57                             & 30                          & 0.29                            & 2.41                              \\ \hline

\end{tabular}
\caption{Table summarizing the different experimental results shown in the main text. For each test, the measured speed $\dLdot$ is reported, together with the measured amplification.}
\label{table:TestResults}
\end{table}

\section{Bifurcation behavior\label{sec:SI:Bifurcation}}

In this section, we describe in more detail the analysis that leads to the bifurcation diagram shown in fig.~1c of the main text.

\subsection{Governing equations and non-dimensionalization}

Consider an initially straight, thin strip of length $L$, thickness $h$ and width $b$ clamped at its two ends with angle $\alpha$ as shown in fig.~\ref{fig:ExpBif}b). We assume that the material is isotropic and homogeneous with Young's modulus $E$ and density $\rho$. Since the strip is slender ($h,b\ll l$), we henceforth assume that it is unshearable, inextensible and ignore any damping; in this limit, small transverse deflections, $\bar{\zeta}(\bar{x},\bar{t})$, must satisfy the dynamic beam equation
\begin{gather}
    \rho bh \frac{\partial^2 \bar{\zeta}}{\partial \bar{t}^2}+B\frac{\partial^4 \bar{\zeta}}{\partial \bar{x}^4}+T \frac{\partial^2 \bar{\zeta}}{\partial \bar{x}^2}=0 \label{eqn:BeamEqnDim}\\
     \int^{\frac{L}{2}}_{-\frac{L}{2}} \left(\frac{\partial \bar{\zeta}}{\partial \bar{x}}\right)^2~d\bar{x}=2\frac{\Delta L}{L}\\
    \bar{\zeta}\left(\pm \frac{L}{2},\bar{t}\right)=0,\quad \bar{\zeta}_x\left(\pm \frac{L}{2},\bar{t}\right)=\mp \alpha,
\end{gather} where $\bar{\zeta}$ is the vertical displacement of the strip, $\alpha$ is the clamping angle, $T$ is then compressive force applied to the strip and $B=\frac{Eh^3b}{12}$ is the bending stiffness. The vertical displacement at the center of the strip is denoted by $h$ in the main text.

To non-dimensionalize the equations, we introduce the following dimensionless variables
\begin{align}
    \zeta = \frac{\bar{\zeta}}{L},\quad x=\frac{\bar{x}}{L},\quad \tau^2=\frac{TL^2}{B},\quad t=\frac{\bar{t}}{t_*},\quad t_*=L^2 \sqrt{\frac{\rho bh}{B}}
\end{align} where we use $\tau^2$ to denote the dimensionless compressive force for later notational convenience. 

Then eqn.~\eqref{eqn:BeamEqnDim} can be re-written in  dimensionless form as
\begin{gather}
   \label{eq:Non1}  \frac{\partial^2 \zeta}{\partial t^2}+\frac{\partial^4 \zeta}{\partial x^4}+\tau^2 \frac{\partial^2 \zeta }{\partial x^2}=0\\
   \label{eq:Non2}  \int^{\frac{1}{2}}_{-\frac{1}{2}} \left(\frac{\partial \zeta}{\partial x}\right)^2~dx=2\frac{\Delta L}{L}\\
  \label{eq:Non3}   \zeta \left(\pm \frac{1}{2},t\right)=0,\quad \zeta_x\left(\pm \frac{1}{2},t\right)=\mp \alpha.
    \end{gather}

\subsection{Static bifurcation behavior}
In equilibrium, eqn.~\eqref{eq:Non1}--\eqref{eq:Non3} can be simplified to 
\begin{gather}
    \label{eq:StaticEq}\frac{d^4 \zeta}{d x^4}+\tau^2 \frac{d^2 \zeta }{d x^2}=0\\
    \label{eq:BCInt}  \int^{\frac{1}{2}}_{-\frac{1}{2}} \left(\frac{d \zeta}{d x}\right)^2 ~dx=2\frac{\Delta L}{L}\\
    \label{eq:BCeq} 
    \zeta \left(\pm \frac{1}{2}\right)=0,\quad \left.\frac{d\zeta}{dx}\right|_{x=\pm \frac{1}{2}}=\mp \alpha.
    \end{gather}
The above equations have the general solutions
\begin{align}
    \label{eq:generalsolu}\zeta=A \sin{\tau x}+B \cos{\tau x}+C+D x
\end{align} where the constants $A,B,C$ and $D$ can be determined from the boundary conditions eqn.~\eqref{eq:BCeq}, which  now read:
\begin{align}
\label{eq:BCeqns}
\begin{split}
 0 & =A\sin{\frac{\tau  }{2}}+B\cos{\frac{\tau}{2}}+C+\frac{D}{2}\\
 0 & =-A\sin{\frac{\tau  }{2}}+B\cos{\frac{\tau}{2}}+C-\frac{D}{2}\\
 \alpha & =A \tau \cos{\frac{\tau}{2} } -B \tau \sin{\frac{\tau}{2} } +D\\
 -\alpha & =A \tau \cos{\frac{\tau}{2} } +B \tau \sin{\frac{\tau}{2} } +D
\end{split}
\end{align}
   
There are two types of solutions to eqn.~\eqref{eq:BCeqns}: symmetric solutions, in which $A=D=0$, and asymmetric solutions ($A,D\neq0$), which are only possible if $\tan{\frac{\tau}{2}}=\frac{\tau}{2}$.

\subsubsection{Symmetric solutions}

For the symmetric case, the solution is 
\begin{align}
 \zeta _{\eqm}^s=\frac{\alpha }{\tau } \left(\frac{ \cos{\tau x} }{\sin{\frac{\tau}{2}}}-\cot\frac{\tau} {2}\right)
 \label{eqn:ZetaSym}
\end{align}
Here $\tau$ is not known a priori, but must be found from the condition of inextensibility eqn.~\eqref{eq:BCInt}, which gives
\begin{align}
  \label{eq:muSym} \mu= \frac{\Delta L }{\alpha ^2 L}=\frac{\tau - \sin{\tau} }{4 \tau  \sin^2{\frac{\tau}{2}} }
\end{align}
we shall notice that the value of $\tau$ is solely dependent on the single parameter $\mu$, which emerges as  the ratio between the relative compression applied and the clamping angle imposed to the strip.

Equations \eqref{eqn:ZetaSym} and \eqref{eq:muSym} allow the bifurcation diagram to be plotted parametrically in terms of $\tau$, for example showing $h=\zeta(0)$ as a function of $\mu$. This is what is plotted in fig.~1c of the main paper.

Fig. ~\ref{fig:ExpBif} depicts $\mu(\tau)$ from \eqref{eq:muSym} with three lowest solutions of $\tau$ by the black, brown and grey curves. The natural mode  corresponds to the lowest value of $\tau$ (black curve) while the inverted mode corresponds to the second lowest value of $\tau$ (brown curve). The plot reveals that there is a local minimum of $\mu$, which is readily calculated to occur at $\tau\approx9.203$ and corresponds to $\mu=\mubif\approx0.247$. For $\mu>\mubif\approx 0.247$, both the natural mode and the inverted mode exists while for  $\mu<\mubif$, only the natural mode exists --- a saddle node bifurcation occurs at $\mu=\mubif$.

\subsubsection{Asymmetric solutions}

Asymmetric solutions exist only if $\tau=\tau_*$ such that $\tau_*/2=\tan(\tau_*/2)$. We find numerically that $\tau_* \approx 8.987$  and the solution may be written
\begin{align}
  \zeta _{\mathrm{eqm}}^a(x)=\frac{\alpha }{\tau _*}\left(\frac{\cos {\tau _* x} }{\sin {\frac{\tau _* }{2}}}-\cot \frac{\tau _* }{2}\right)+\frac{\alpha }{\tau _*} D \left( \sin{\tau _* x}-2 x \sin{ \frac{\tau _*}{2}}  \right). 
  \label{eqn:ShapeAsym}
\end{align} Note that the symmetric portion of \eqref{eqn:ShapeAsym} is precisely that of \eqref{eqn:ZetaSym}, albeit with $\tau$ replaced by $\tau_*$. The size of the asymmetry, encoded by $D$, is determined from the condition of inextensibility eqn.~\eqref{eq:BCInt}; we find that
\begin{align}
    D=\frac{\sqrt{\frac{4 \Delta L }{L\alpha ^2}-1}}{\sin{\frac{\tau_*}{2}}}.
\end{align} Note that this constant is only real when $\Delta L/(L \alpha ^2)=\mu> 1/4$ --- this solution vanishes for $\mu\leq1/4$.

The bifurcation diagram fig.~\ref{fig:ExpBif}c) is completed by adding the asymmetric mode (cyan curve), noting that the mid-point displacement $\zeta(0)=\frac{\alpha }{\tau _*}\left(\frac{1-\cos\frac{\tau _*} {2}}{\sin\frac{\tau _* }{2}}\right)$ is constant, independent of $\mu\geq1/4$.

\subsection{Stability analysis}

Having determined the various equilibrium solutions that exist at a given $\mu$, we now determine their stability. The stability of a given equilibrium is determined by small perturbations. For an equilibrium solution $\zeta_{\eqm}(x)$, we perturb it and assume a solution of the form
\begin{gather}
    \label{eqn:perturbExpression}\zeta_n(x,t)=\zeta_{\eqm}(x)+\epsilon \zeta_p(x) e^{\lambda t}\\
    \tau_n(t)=\tau_{\eqm}+\epsilon \tau_p e^{\lambda t}
\end{gather} for some unknown growth rate $\lambda$, and unknown perturbation quantities $\zeta_p(x)$ and $\tau_p$.

Substituting this solution back into eqn.~\eqref{eq:Non1}-\eqref{eq:Non3} and considering only terms of $O(\epsilon)$, we obtain
\begin{gather}
    \frac{d^4\zeta_p}{d x^4}+\tau_{\eqm}^2 \frac{d^2\zeta_p}{d x^2}+\lambda^2 \zeta_p+2\tau_p \tau_{\eqm} \frac{d ^2 \zeta_{\eqm}}{d x^2}=0\label{eqn:PertubEqn}\\
    \int_0^1 \frac{d\zeta_{\eqm}}{d x}\frac{d\zeta_{p}}{d x}dx=0\\
    \zeta_p(0,t)=\zeta_p(1,t)=\frac{d \zeta_p}{d x}(0,t)=\frac{d \zeta_p}{d x}(1,t)=0\label{eqn:PertubEqnBCs}
\end{gather}
The general solution of this system can be expressed as
\begin{gather}
    \label{eqn:perturbSoluGeneral}
        \zeta_p=A \cos{\mu_+ x}+B \sin{\mu_+ x}+ C\cos{\mu_- x}+ D \sin{\mu_- x}-2 \tau_p\tau_{\eqm} \frac{\partial^2 \zeta_{\eqm}}{\partial x^2} \frac{1}{\lambda^2}\\
    \label{eqn:perturbSoluGeneral2}
    \mu_{\pm}=\left(\frac{\tau_{\eqm}^2 \pm\sqrt{{\tau_{\eqm}^4-4\lambda^2}}}{2}\right)^{1/2}.
\end{gather}
Any solution must also satisfy the boundary conditions eqn.~\eqref{eqn:PertubEqnBCs}.  
Substituting the expressions for $\zeta_{\eqm}$ and $\tau_{\eqm}$ into eqn.~\eqref{eqn:perturbSoluGeneral} and \eqref{eqn:PertubEqnBCs} we obtain a system of five linear equations in five unknowns ($A,B, C, D$ and $\tau_p$). We can then find (numerically) the values of $\lambda^2$ for which non-trivial solutions of this homogeneous linear system exist. 

Recalling from eqn.~\eqref{eqn:perturbExpression} that the perturbation to the equilibrium state grows with time like $e^{\lambda t}$, we note that unstable solutions correspond to  $\lambda^2>0$ while stable solutions correspond to $\lambda^2<0$, (i.e.~undamped oscillations). We use the sign of $\lambda^2$ to determine the stability of the different equilibrium solutions already determined. The results of the stability analysis are plotted in fig.~\ref{fig:ExpBif}c where the dashed line indicates unstable mode and the solid line indicates that the equilibrium is stable.

\subsection{Natural Frequency of oscillations} \label{sec:NaturalFrequencyArch}

The linear stability analysis also allows us to find the period of
oscillations about the (stable) natural state. The time scale that arises from this analysis is important because, as we shall discuss later in \S \ref{sec:TimeScaling}, it provides a natural time scale in the continuous arch model  that can be compared with the simple toy model.  

\begin{figure}[t]
\includegraphics[width=0.6\columnwidth]{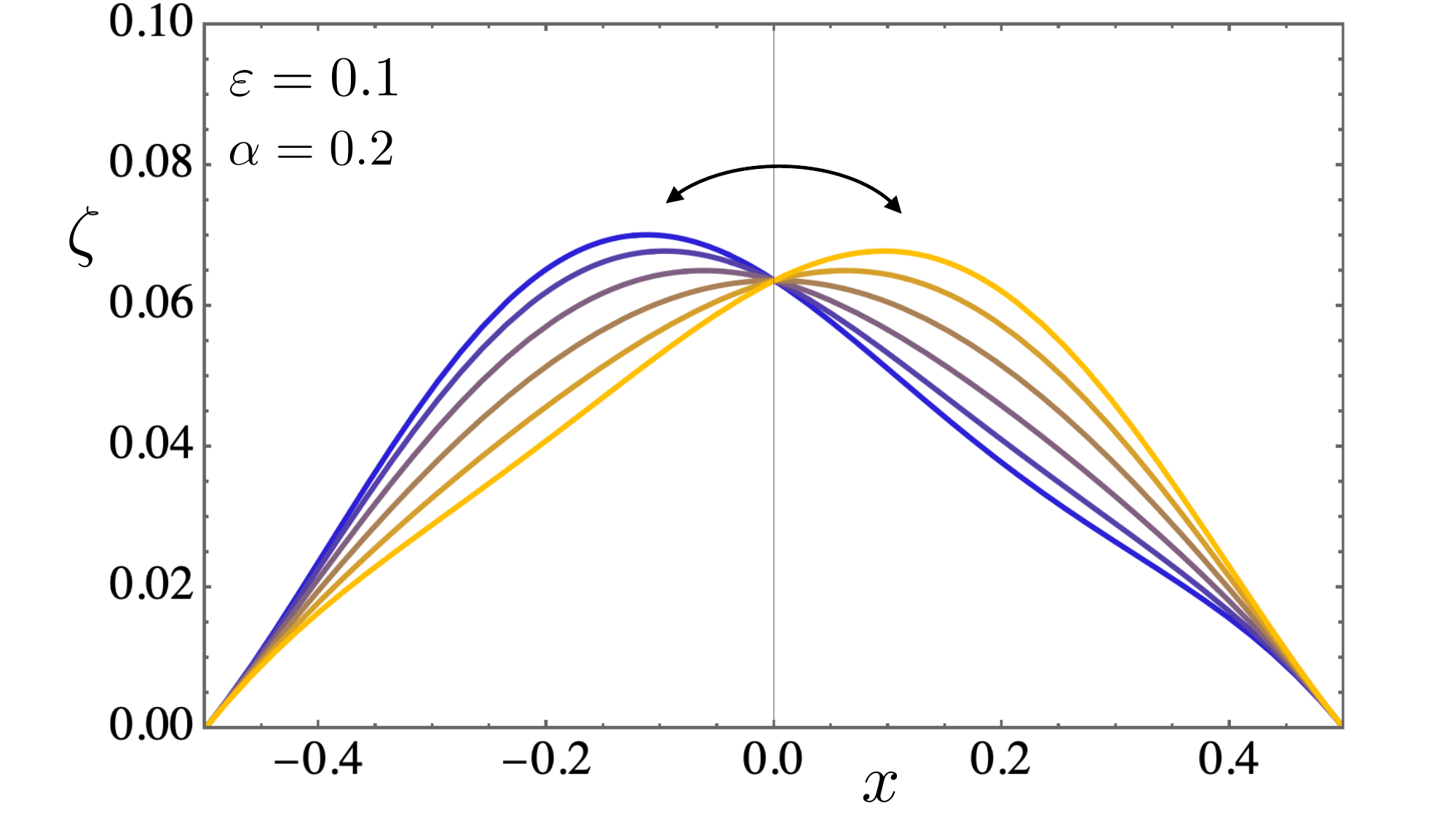}
\caption{The asymmetric mode of oscillation of the strip in the natural state at $\mu=0.25$. }
\label{fig:oscillationsarch}
\end{figure}

The period of frequency varies slightly as a function of the end-shortening, $\mu$. Therefore, we choose to evaluate it at the end-stretching at which the arch first goes unstable, $\mu=0.25$. This occurs at the value of $\tau_{\mathrm{eqm}}=\pi$. 
Substituting these values into \eqref{eqn:perturbSoluGeneral} and \eqref{eqn:perturbSoluGeneral2} and solving the problem numerically reveals that the oscillations correspond to horizontal oscillations of the arch, as shown in figure \ref{fig:oscillationsarch} and found in a related problem \cite{Pandey2014}; moreover, the dimensionless frequency $\omega =\sqrt{-\lambda^2}=\sqrt{-\lambda^2}\approx\sqrt{3348.04}\approx57.86$. Note that this is a slightly larger value than that reported in \cite{Pandey2014} for the case of a horizontally clamped strip ($\alpha=0$).
This oscillation introduces a natural (dimensionless) time period  $T_{\mathrm{osc}}=2 \pi/\omega=0.108$; in the main text, we suggest that this period is a natural time scale with which to rescale time.

\section{Numerical simulations\label{sec:SI:Nums}}

In this section, we discuss the numerical simulation method used to study the snap-through dynamics of elastic arches. Our method is based on the Discrete Elastic Rods algorithm  \cite{bergou2008discrete,bergou2010discrete} and more details are given below. The physical and geometric parameters used in the numerical simulation are as follows: beam length $L = 108$ mm, beam width $w = 25.0$ mm, beam thickness $h = 0.0508$ mm, Young's modulus $E=200$ GPa, material density $\rho=8000$ kg/m$^3$, initial compress $\Delta L / L = 0.2$, and rotational angle $\alpha = \pi / 12$. (The Poisson ratio $\nu$ does not enter our calculations because the strip is narrow and so the appropriate moment of inertia is $I=h^3w/12$, independent of $\nu$ \cite{Audoly2010}.)

The configuration of a planar elastic strip can be described based on its centerline, $s$. From the numerical point of view, we discretize the strip centerline into $ N $ nodes so that the  degrees of freedom are
\begin{equation}
\mathbf{q} \equiv \left[ \mathbf{x}_{0}, \mathbf{x}_{1}, ..., \mathbf{x}_{N-1} \right] \in \mathbb
{R}^{2N \times 1},
\end{equation}
and the $i$-th edge vector is given by,
\begin{equation}
\mathbf{e}_{i} = \mathbf{x}_{i+1} - \mathbf{x}_{i}, \; \mathrm{with} \; i \in [0, \ldots, N-2].
\end{equation} 

The energy of a strip of linearly elastic material is broken down into its stretching and bending energies. For a strip of length $L$, cross-sectional area $w \times h$, and Young's modulus $E$, the discrete stretching energy of the edge connecting $\mathbf x_i$ and $\mathbf x_{i+1}$ is 
\begin{equation}
E^s_i = \frac{1}{2} \; EA \; \varepsilon_{i}^2 \; || \bar{\mathbf{e}}_{i} ||,
\end{equation}
where $EA$ is the stretching stiffness, $|| \bar{\mathbf{e}}^i ||$ is the undeformed length of $i$-th edge vector (quantities with a bar on top indicate that quantity in its undeformed configuration), and $\varepsilon_{i}$ is its axial stretch strain,
\begin{equation}
\varepsilon_{i} = \frac { ||{\mathbf{e}}_{i} || } { || \bar{\mathbf{e}}_{i} || } - 1.
\end{equation}
The discrete bending energy is associated with discrete curvature, and can be written
\begin{equation}
E^b_i = \frac{1}{2} \; EI \; \left( \kappa_i - \bar{\kappa}_i \right)^{2} \Delta \bar{L}_{i}
\end{equation}
where $EI$ is the bending stiffness, $\Delta \bar{L}_{i} =  { (||{ \bar{\mathbf{e}}}_{i-1} || + ||{ \bar{\mathbf{e}}}_{i} ||)} / {2}$ is the mean length of the $(i-1)$-st and $i$-th edges.
Moreover, $\kappa_i$ is the discrete curvature,
\begin{equation}
\kappa_i = \frac { 2 \tan(  {\phi_i} / {2} ) } { \Delta \bar{L}_{i}},
\end{equation}
and $\phi_{i}$ is the turning angle between the two consecutive edges, $\{\mathbf{e}_{i-1}, \mathbf{e}_{i} \}$.

The total stretching energy of the elastic strip system can be obtained simply by summing over all the nodes and edges, i.e., 
\begin{equation}
E^s = \sum_{i=0}^{N-2} E^s_i \; \; \mathrm{and} \; \; E^b = \sum_{i=1}^{N-2} E^b_i.
\end{equation}
It is worth noting that the strip is effectively inextensible during the snap-through process: the elastic stretching energy penalty is so large that in practice the uniaxial strain is zero, i.e.~$\varepsilon \equiv 0$. 

We numerically solve the discrete equations of motion step by step by using a second-order symplectic Newmark-beta method \cite{huang2019newmark}.
At the $k$-th time step, $t_{k}$, the generalized displacement vector $\mathbf{q}$ can be updated from $t=t_{k}$ to $t=t_{k+1}$ by imposing the equation of motion, as
\begin{equation}
\begin{aligned}
\mathbb{M} \left( \Delta \mathbf{q}_{k+1} - \dot{\mathbf{q}}_{k} \;  \delta t \right) &- \frac {1} {4} \; \delta t ^2 \; \left( \mathbf{F}^{\text{int}}_{k+1} + \mathbf{F}_{k+1}^{\text{ext}} \right) - \frac {1} {4} \; \delta t ^2 \; \left( \mathbf{F}^{\text{int}}_{k} + \mathbf{F}_{k}^{\text{ext}} \right) = \mathbf{0} \\ 
\mathbf{q}_{k+1} &= \mathbf{q}_{k} + \Delta \mathbf{q}_{k+1} \\
\dot{\mathbf{q}}_{k+1} &= \frac {2} {h } \Delta \mathbf{q}_{k+1} - \dot{\mathbf{q}}_{k},
\end{aligned}
\label{eq:NewmarkBeta}
\end{equation}
where $\mathbb{M} $ is the diagonal mass matrix, $\delta t $ is the time step size, $\mathbf{F}^{\text{int}} = - \nabla (E^{s} + E^{b})$ is the internal elastic force vector, and $\mathbf{F}^{\text{ext}}$ is the external force vector (e.g., gravity and damping).
The subscript indicates the evaluation at $t=t_{k}$ (or $t=t_{k+1}$) time step.
The classical Newton-Raphson method is applied to iteratively solve the nonlinear equations of motion.
While the error remains larger than the tolerance, we repeat the iteration process, stopping only once the error falls below the tolerance and then moving on to the next time step.

\begin{figure}[t]
\includegraphics[width=\columnwidth]{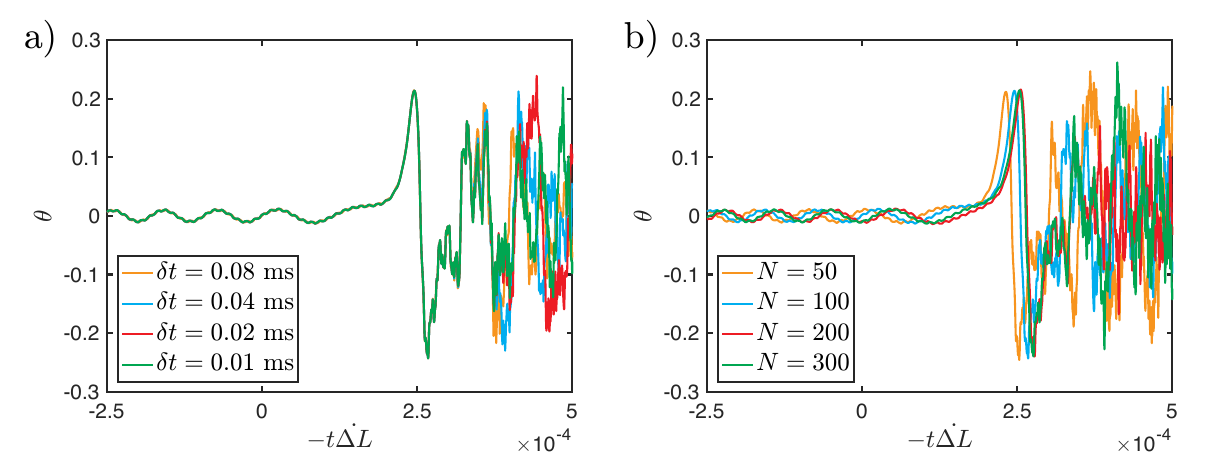}
\caption{Convergence study for both the temporal and spatial discretizations. (a) Central angle $\theta$ as a function of boundary stretching distance $- t \dot{\Delta {L}}$, with different time steps: $\delta t \in \{0.08, 0.04, 0.02 ,0.01 \}$ ms and a fixed number of nodes ($N = 100$). (b) Central angle $\theta$ as a function of boundary stretching distance $ - t \dot{\Delta {L}}$, with a different fixed number of nodes ($N \in \{50, 100, 200, 300 \}$) and a fixed time step size $ \delta t = 0.01$ ms.}
\label{fig:convergentPlot}
\end{figure}

To achieve a clamped--clamped boundary condition, the first two nodes, $\{\mathbf{x}_{0}, \mathbf{x}_{1} \}$, as well as the last two nodes, $\{\mathbf{x}_{N-2}, \mathbf{x}_{N-1} \}$, are constrained based on the boundary and loading conditions; all other nodes are free to evolve based on the  dynamic governing equations.
The geometric and physical parameters are identical to the experimental setup. A convergence study for both the spatial and temporal discretizations is presented in Fig.~\ref{fig:convergentPlot}, demonstrating that $N=100$ grid points with $\delta t=0.01\mathrm{~ms}$ is sufficient. 
It is worth noting that a small phase difference can be found in Fig.~\ref{fig:convergentPlot}B. This is caused by the implementation of the clamped boundary conditions, which means that the effective length of the beam is slightly different for different numbers of nodes, $N$. As a result, the natural frequencies are also slightly different: an effect that is manifested as a small variation in the phase of oscillations as time goes by.
Nevertheless, the overall dynamic processes (e.g., amplification) show no variations as $N$ varies.

\subsection{Origin of oscillations}

To attain a desired stretching speed, $\dLdot$,  different acceleration protocols can be used. Our simulations show that these different protocols have an important effect on the magnitude of asymmetric oscillations that are introduced into the system.
\paragraph{Symmetrically moved ends} When the two ends of the arch are moved  simultaneously in opposite directions, there is no oscillation in the angle measured at the beam midpoint, regardless of the size of the acceleration.

\paragraph{Accelerating one end only} If only one end is accelerated, we observe significant asymmetric  oscillations at the beam's midpoint. The magnitude of this  oscillation is related to the loading acceleration: a larger initial acceleration leads to  larger oscillations.  We use different accelerations to control the size of the initial asymmetric  oscillations in the numerical simulation results reported in Fig.~2B,C of the main text. These show that while the amplitude of oscillations varies with acceleration, the amplification is well-defined and hence a natural means to study the onset of visible asymmetry in experiments. These results also suggest that the asymmetric oscillations observed in experiments is likely due to the asymmetric acceleration of the ends, which may be accentuated by non-ideal behavior of the motor (e.g.~non-constant acceleration, or `jerk').

\paragraph{Point force}  To achieve large oscillations, a moderate transverse force is used as a perturbation, i.e.~we first applied a moderate transverse force onto the beam system, then remove it, so that moderate transverse oscillations of the beam are excited. The amplitude of these directly-induced oscillations is related to the magnitude of the perturbation force.

\section{The Toy model \label{sec:SI:Toy}}

Characterizing the behavior of the Elastica's snap-through analytically is complex because the dynamic equations involve both spatial and temporal dependence: we would need to solve for the continuous shape of the arch  at any instant in time. To make analytical progress, we therefore choose to study a simpler system, the double-mass von Mises truss (see Fig.~3a of the main text). The properties of this system can be tuned so that it has the same bifurcation behavior as the equilibrium Elastica (see Fig.~1c of the main text), though its shape is fully described by the position of the two elements (or masses). This model is much simpler to treat mathematically, but retains sufficient degrees of freedom to capture both symmetric and asymmetric transitions dynamically. This allows us to understand the key physics that governs the amplification of the asymmetric mode.

\subsection{The double-mass von Mises truss Model}

\begin{figure}[h]
    \centering
    \includegraphics[width=\textwidth]{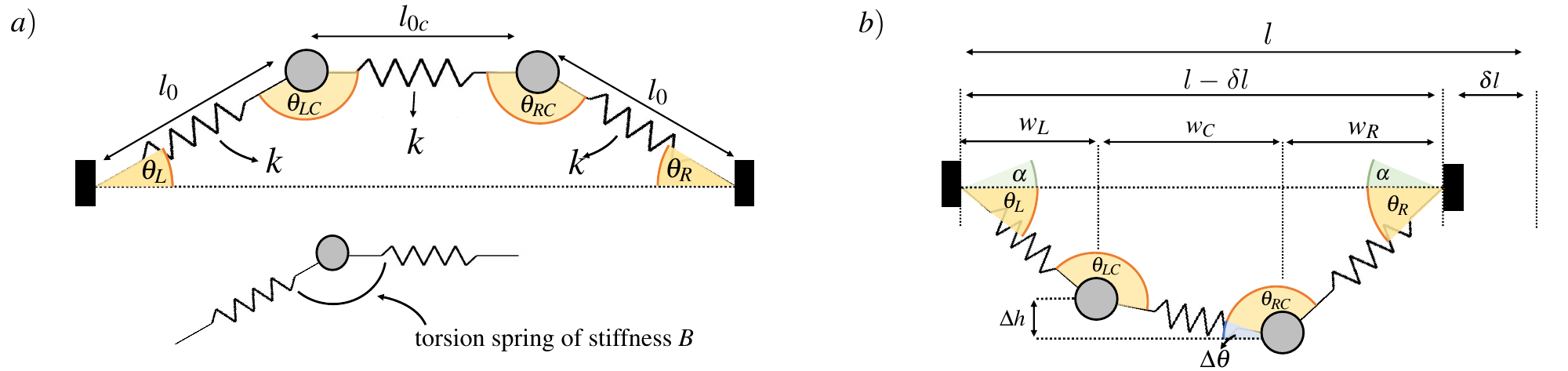}
    \caption{a) Sketch of the relaxed double-mass von Mises truss with two elements (masses) and torsion springs acting about the angles $\theta_L$, $\theta_R$,$\theta_{LC}$ and $\theta_{RC}$. In this relaxed states, the linear springs take their natural lengths, that is $l_0$ for the edge springs and $l_{0c}$ for the edge springs. b) An example of the inverted state of the double-mass von Mises truss after compression by $\delta l$. }
    \label{fig:doublemise}
\end{figure}

The double-mass von Mises truss model discussed in the main text is shown in detail in fig.~\ref{fig:doublemise}. It is designed in the following way: two springs of natural length $l_0$ and stiffness $k$ connect two point-particles, each of mass $m$. We refer to these as the left or right mass and spring and use a subscript $L$ or $R$ to identify them. The two masses are connected by a third spring of the same stiffness $k$ and natural length $l_{0C}$. 

All springs are connected to the mass and to each other via torsional  springs of stiffness $B$. The torsional springs that connect the left and right springs to the edges of the structure are relaxed when the spring makes an angle $\alpha$ below the horizontal. This condition mimics the boundary condition on the Elastica, in which its tangent is clamped at an angle $\alpha$ above the horizontal. 
Conversely, the torsion springs connecting the left and right springs to the central one are relaxed when the springs are aligned or straight (i.e.~when the angle between them is $\pi$) providing a bending stiffness to the system, analogous to that of an Elastica that is naturally flat. We therefore refer to $B$ as the bending stiffness. 
Without the bending stiffness provided by the second set of torsion springs, the system is always unstable to asymmetric deformations. Since we want to capture the qualitative behavior of the Elastica, we need the system to be stable in the inverted state for some values of the compression $\delta l$ and the torsion springs are therefore an essential ingredient in the model.

\section{Static analysis of the double-mass von Mises truss}

We begin by studying the stability of the inverted state of the double-mass von Mises truss and determine its bifurcation diagram to gain insight into its behavior. Later, we shall address the dynamics of the system and study how the unloading rate affects the snap-through instability.

\subsection{The energy of the double truss}

In this section, we aim to write down the static (or potential) energy of the system and study the stability of the inverted state. 
As shown in fig.~\ref{fig:doublemise}, we define the reference length of the system $l=2l_0+l_{0C}$ to be the sum of the relaxed lengths of each spring. We also assume the end distance can be changed by an amount $\delta l$, causing a deformation. In the deformed state, we let $l_R$ and $l_L$ be the length of the left and right springs respectively,  $w_L$ and $w_R$ the horizontal distance of the left mass and right mass from their respective edges, $\Delta h$ the difference in height between the two masses, and $l_C$ and $w_C$ be the length of the central spring and its horizontal width. Moreover, we let $\theta_L$ and let $\theta_R$ be the angles the side springs make with the horizontal while $\theta_{LC}$ and $\theta_{RC}$ are the angles between the side springs and the central one, with $\Delta \theta$ being the angle the central spring makes with the horizontal, as shown schematically in figure \ref{fig:doublemise}b). 
Using elementary geometry, we have that:
\begin{align}
    l_L&=\frac{w_L}{\cos\theta_L},\\
    l_R&=\frac{w_R}{\cos\theta_R},\\
    w_C&=l-\delta l-w_L-w_R,\\
    \Delta h=&w_L \tan\theta_L-w_R \tan\theta_R,\\
    l_C&=\sqrt{\Delta h^2+w_c^2},\\
    \Delta \theta&=\arctan(w_c,\Delta h),\\
    \theta_{LC}&=(\pi/2-\theta_{L})+(\pi/2-\Delta \theta),\\
    \theta_{RC}&=(\pi/2-\theta_{R})+(\pi/2+\Delta \theta).
\end{align}
We can use these relations to write the energy stored in the linear springs as:
\begin{equation}
    U_{springs}=\half k\left[(l_L-l_0)^2+ (l_R-l_0)^2+ (l_C-l_{0C})^2 \right]
\end{equation}
and the bending energy stored in the torsion springs as
\begin{equation}
    U_{B}=\half B\left[(\theta_L-\alpha)^2+(\theta_R-\alpha)^2 +(\theta_{LC}-\pi)^2+(\theta_{RC}-\pi)^2\right].
\end{equation}
The total energy of the system is $U=U_{springs}+U_B$ and can be written as a function of the variables $w_L$, $w_R$, $\theta_L$, $\theta_R$ as well as the control parameter $\delta l$:
\begin{align}
\notag
 U=&\frac{1}{2} k \left[l_{0C}-\sqrt{\left(l- \delta l- w_L- w_R\right)^2+\left(w_L \tan \theta_L-w_R \tan \theta_R \right)^2}\right]^2\\
  \notag
    &+\frac{1}{2} k\left(\left(l_0-w_L \sec \theta_L\right)^2+\left(l_0-w_R \sec \theta _R\right){}^2\right)\\ 
    \notag
  &+\frac{1}{2} B \left[\arctan \left(l-\delta l-w_L-w_R,w_L \tan \theta_L-w_R \tan \theta_R \right)+\theta_L\right]^2\\
    \notag
    &+\frac{1}{2} B\left[\arctan \left(l-\delta l-w_L-w_R,w_L \tan \theta_L-w_R \tan \theta _R\right)-\theta _R\right]^2\\
   &+\frac{1}{2} B \left[\left(\alpha-\theta_L\right)^2+\left(\alpha -\theta_R\right)^2\right].
   \label{eqn:totalenergy}
\end{align}
(Note that here $\arctan(x,y)$ is the two-argument arctangent; that is, for $x$, $y$ in the correct quadrant, $\arctan(x,y)=\arctan(y/x)$.)
To make progress, we begin by using a suitable change of variable to emphasize the asymmetric behavior. In particular, we introduce the symmetric and asymmetric parts of the angle:
 $$\theta_S=\half (\theta_L+\theta_R)$$ and $$\theta_A=\theta_L- \theta_R,$$ respectively. Similarly, we can decompose the change in width into symmetric, $w_S=w_L+w_R- (2 l_0-\delta l)$, and asymmetric, $w_A=w_R-w_L$, parts. In this way, we may write both the angles and change in width as:
\begin{align}
    \theta_L&=\theta_S+\half \theta_A\\
    \theta_R&=\theta_S-\half \theta_A\\
    w_L&=l_0-\half(\delta l-w_S+w_A)\\
    w_R&=l_0-\half(\delta l-w_S-w_A).
\end{align}

Our primary focus in the main text is on the evolution of the angles $\theta_S$ and $\theta_A$. To simplify this analysis and make analytical progress, we focus now on the behavior of the energy for small values of the clamp angle $\alpha$.

\subsection{Energy expansion for small  $\alpha$}

For small  $\alpha$, both the natural and inverted configurations are, in some sense, close to the flat case (in which the masses are aligned with the edges of the structure). This is equivalent to the small vertical displacement and small gradient approximations used to determine the bifurcation behavior of the continuous Elastica problem in \S\ref{sec:SI:Bifurcation} and is hence also consistent with the small values of $\alpha$ used in experiments.

We observe that for the system to be almost flat, we require the angles $\theta_R$, $\theta_L$, as well as $\Delta \theta$, to be small. Clearly, $\theta_R$, $\theta_L$ will be of the same size as $\alpha$. Thus, the angle $\alpha$ determines how close we are to the flat case and is our expansion parameter.

To understand how  quantities such as $w_R$ and $w_L$ scale with $\alpha$, let us consider the unconstrained system in which the length $l$ is free, $\alpha$ is small, and the system can relax. In this case,  $\theta_i \sim \alpha$ for $i=L,R$. If we do not constrain the end-to-end displacement, the springs will be relaxed so that $l_i=l_0$ and the width changes according to $w_i \sim l_0 \cos \theta_i \sim l_0(1 - \half \alpha^2)$, meaning that the change in horizontal length scales like $\sim \alpha^2$. We can thus rescale angles by $\alpha$ and changes in horizontal lengths by $\alpha^2$.

We summarize our rescalings as follows:
\begin{align}
\notag
    \theta_L&=\alpha (\psi_S+\half \psi_A)\\
    \notag
    \theta_R&=\alpha(\psi_S-\half \psi_A)\\
    \notag
    w_L&=l_0\bigl[1-\half \alpha^2 (\mu - W_S+W_A)\bigr]\\
    \notag
    w_R&=l_0\bigl[1-\half \alpha^2 (\mu - W_S-W_A)\bigr]\\
    \notag
    l_{0C}&=l_0 L_{0C}\\
    B&=\third \beta \alpha^2 k l_0^2
    \label{eqn:rescalings}
\end{align}
where $\mu=\delta l/(\alpha^2 l)$, $\psi_A=\theta_A/\alpha$, $\psi_S=\theta_S/\alpha$, $W_S=w_S/(l_0 \alpha^2)$, $W_A=w_A/(l_0 \alpha^2)$.  (The bifurcation parameter $\mu=\delta l/(\alpha^2 l)$ has emerged again, as in the continuous case.)

Note that the energy of the springs scales like the change in length squared, meaning $U_{spring} \sim k (l_0 \alpha^2)^2$, while the bending energy scales like the change in angle squared $U_B \sim B \alpha^2$. For the two to balance, we need $B \sim \beta \alpha^2 k l_0^2$ (the factor of one-third being included to simplify later equations). 

We can now expand the energy to leading order ($\alpha^4$). The rescaled energy $\tilde{U}=\frac{U}{kl_0^2 \alpha^4}$ is then given by:
\begin{align}
\notag
192\tilde{U}=&\frac{8 \psi _A^2 (4 \beta  (L_{0C}
   (L_{0C}+2)+2)-3 \mu  L_{0C} (5
   L_{0C}+4))}{L_{0C}^2}+\frac{3
   \left(L_{0C}^2+8\right) \psi
   _A^4}{L_{0C}^2}+48 \psi _S^4\\
   \notag
   &+\frac{24 W_S
   \left((L_{0C}-4) \psi _A^2+4
   L_{0C} \left(\mu +2 \mu 
   L_{0C}+\psi
   _S^2\right)\right)}{L_{0C}}+8 \psi
   _S^2 \left(9 \psi _A^2+16 \beta -12 \mu
   \right)\\
   &-96 W_A \psi _A \psi _S+48
   W_A^2+64 \beta +48 \mu ^2 (2 L_{0C}
   (L_{0C}+2)+3)-128 \beta  \psi _S+144
   W_S^2
\label{eqn:energy2}
\end{align}
This energy is easily minimized with respect to variations in $W_S$ and $W_A$ by solving $\p{U}{W_A}=0$ and $\p{U}{W_S}=0$, leading to:
\begin{align}
\label{eqn:WS}
    W_S&= -\frac{(L_{0C}-4) \psi _A^2+4 L_{0C}
   \left(\mu(1 +2 L_{0C})+\psi
   _S^2\right)}{12 L_{0C}},\\
   \label{eqn:WA}
   W_A &= \psi _A \psi _S.
\end{align}
Substituting these results back into eqn.~\eqref{eqn:energy2} we obtain the rescaled energy:
    \begin{align}
\tilde{U}=U_0-a_1 \psi_S+\half (a_2-a_3 \mu)\psi_S^2+\fourth a_4 \psi_S^4+\half (b_1 -b_2 \mu+b_3 \psi_S^2)\psi_A^2+\fourth b_4 \psi_A^4,
\label{eqn:simpleenergy}
\end{align}
where:
\begin{align}
    U_0&= \frac{1}{6}\left( 2 \beta+(2+L_{0C})^2 \mu^2 \right)\\
    a_1&= \frac{2}{3} \beta,\quad a_2= \frac{4}{3} \beta,\quad    a_3= 2(2+L_{0C})/3,\quad a_4=\frac{2}{3},\\
    b_1&= \frac{2+L_{0C}(2+L_{0C})}{3 L_{0c}^2}\beta\\
    b_2&= \frac{(2+L_{0C})^2}{6 L_{0C}},\quad     b_3=\frac{2+L_{0C}}{6 L_{0C}},\\
    b_4&=\frac{(2+L_{0C})^2}{24 L_{0C}^2}.
\end{align}
Note that the signs of terms in \eqref{eqn:simpleenergy} have been chosen so that all of the above constants are positive.

The energy of \eqref{eqn:simpleenergy} is precisely the one that appears in equation (3) of the main text, where we have collected the prefactor of the $\psi_A^2$ term and introduced
\begin{equation}
    f(\psi_S;\mu):=b_1 -b_2 \mu+b_3 \psi_S^2.
    \label{functionf}
\end{equation}
The advantage of writing the energy in this form is that its structure becomes simple and compact, allowing us to study the equilibria of the system. From now on, perhaps confusingly, we shall write some results with respect to the constant defined above, and others explicitly as functions of the parameters in the problem. This is because sometimes the results are too complex to write explicitly as functions of the parameters, while in other cases they are simple enough allowing us to spell them out without having to use the constants.

\subsection{The energy minima and their stability}

The energy in equation \eqref{eqn:simpleenergy} is a simple quartic polynomial in both $\psi_S$ and $\psi_A$. To study the equilibria of the system at a given value of $\mu$, we differentiate \eqref{eqn:simpleenergy} with respect to $\psi_S$ and
 $\psi_A$ and simultaneously solve 
 \begin{equation}
 \label{eqn:equilibria}
     \frac{\partial \tilde{U}}{\partial \psi_S}=\frac{\partial \tilde{U}}{\partial \psi_A}=0.
 \end{equation}

\subsubsection{Symmetric equilibria: $\psi_A=0$}
 
It is immediately apparent that $\psi_A=0$ is one solution of the second equilibrium equation (since $\tilde{U}$ contains only even powers of $\psi_A$). To find the associated value of $\psi_S$ in this symmetric state, we must solve:
 \begin{align}
     \frac{\partial \tilde{U}}{\partial \psi_S}\bigg|_{\psi_A=0}=-a_1+(a_2-a_3 \mu)\psi_S+a_4 \psi_S^3=0.
 \end{align}
This equation has three zeroes (equilibria) which may not  be real, depending on the value of $\mu$. They  are given by:
\begin{align}
\label{eqn:EquilibriaSimple}
    \psi _N= & \frac{\sqrt[3]{2} \left(a_3 \mu-a_2\right)}{\sqrt[3]{\sqrt{108 a_4^3
   \left(a_2-a_3 \mu\right){}^3+729 a_1^2 a_4^4}+27 a_1
   a_4^2}}\\
   &+\frac{\sqrt[3]{\sqrt{108 a_4^3 \left(a_2-a_3 \mu\right){}^3+729 a_1^2
   a_4^4}+27 a_1 a_4^2}}{3 \sqrt[3]{2} a_4}\\
   \notag
   \psi _I= &\frac{\left(1+i \sqrt{3}\right) \left(a_2-a_3 \mu\right)}{2^{2/3}
   \sqrt[3]{\sqrt{108 a_4^3 \left(a_2-a_3 \mu\right){}^3+729 a_1^2 a_4^4}+27 a_1
   a_4^2}}\\
   \notag
   &-\frac{\left(1-i \sqrt{3}\right) \sqrt[3]{\sqrt{108 a_4^3 \left(a_2-a_3
   \mu\right){}^3+729 a_1^2 a_4^4}+27 a_1 a_4^2}}{6 \sqrt[3]{2} a_4}\\
   \psi _U&=\frac{\left(1-i \sqrt{3}\right) \left(a_2-a_3 \mu\right)}{2^{2/3}
   \sqrt[3]{\sqrt{108 a_4^3 \left(a_2-a_3 \mu\right){}^3+729 a_1^2 a_4^4}+27 a_1
   a_4^2}}\\
   &-\frac{\left(1+i \sqrt{3}\right) \sqrt[3]{\sqrt{108 a_4^3 \left(a_2-a_3
   \mu\right){}^3+729 a_1^2 a_4^4}+27 a_1 a_4^2}}{6 \sqrt[3]{2} a_4}
\end{align}

These solutions correspond to the natural state $\psi_N$, the inverted state $\psi_I$ and an unstable state in between the two, $\psi_U$. The inverted and middle state annihilate when 
$$\frac{d\tilde{U}}{d \psi_S}\bigg|_{\psi_A=0}=\frac{d^2\tilde{U}}{d \psi_S^2}\bigg|_{\psi_A=0}=0,$$
corresponding to a \textbf{saddle-node bifurcation} at the critical values
\begin{align}
    \psi^*_S=-\left(\frac{a_1}{2 a_4}\right)^{1/3}=-\left(\frac{\beta}{2}\right)^{1/3}\,\,\,\,\,\,,\,\,\,\,\,\mubif^{\mathrm{vM}}=\frac{a_2}{a_3}+\frac{3 }{2 a_3}\left(2 a_1^2 a_4\right)^{1/3}=\frac{3 {\psi_S^*}^2+2 \beta}{(2+L_{0C})}.
    \label{eqn:saddlenode}
\end{align}
This means that when $\mu<\mubif^{\mathrm{vM}}$, only one symmetric equilibrium exists, the natural state $\psi_N$.

\subsubsection{Asymmetric equilibria: $\psi_A\neq0$}

There are two other equilibrium solutions with $\psi_A \neq 0$. Again, since $\Tilde{U}$ involves only even powers of $\psi_A$,  we note that if $(\psi_S,\psi_A)=(a,b)$ is an equilibrium, then so is the point $(\psi_S,\psi_A)=(a,-b)$. 
These other equilibrium points are found by solving eqns \eqref{eqn:equilibria} and are best written explicitly as functions of the parameters: 
\begin{align}
\label{eqn:psiSA}
\psi_S&= -\frac{1}{4} L_{0C} \left(L_{0C}+2\right)\\
\psi_A&=\pm\sqrt{\frac{16 L_{0C} (2+L_{0C}) \mu - 32(2+L_{0C}(2+L_{0C})\beta - L_{0C}^3 (2-L_{0C})^3}{2(2+L_{0C})}}
\end{align}
When the two asymmetric (saddle) equilibria meet in a pitchfork bifurcation, locally the system ceases to be stable to asymmetric perturbations. Note that this corresponds to the point where $f(\psi_S;\mu)$ changes from being positive to being negative. This occurs at a critical value, $\mu=\mu_A^{\mathrm{vM}}$, that can be calculated by solving $\half(b_1+b_2 \mu+b_3 \psi_I^2)=0$, telling us that the pitchfork bifurcation occurs at $\psi _S= -\frac{1}{4} L_{0C} \left(L_{0C}+2\right)$ and at the critical value of
\begin{align}
\mu_A^{\mathrm{vM}}=\frac{1}{16} L_{0C}^2
   \left(2+ L_{0C}\right)^2+
   \frac{2(2+L_{0C}(2+L_{0C})\beta}{L_{0C}(2+L_{0C})^2}.
\label{eqn:pitchfork}
\end{align}

\subsubsection{Reconstructing the elastica's bifurcation diagram}

Our static analysis has revealed that the structure of the energy, and thus the behavior of the double-mass von Mises truss, is entirely determined by the choice of three dimensionless parameters. These parameters are $\beta$, indicating the ratio between bending and stretching stiffness, $L_{0C}$, representing the length ratio between the side and central springs, and $\mu$, which characterizes the compression in the system. The latter parameter can be easily controlled and is used to induce the snap-through in the system. Conversely, the first two are material parameters and are kept fixed.

\subsection{Stability of the inverted state} 
If we consider the system in the inverted state with $\psi_A=0$ and imagine slowly decreasing $\mu$ from some large value, we may then ask the question: \emph{when and how does the system go unstable?} To answer, we begin by only considering the quasi-static problem and thus ignore dynamic effects. 

At first, the system is in equilibrium, with $\psi_S=\psi_I$. However, at $\mu=\mu_A$, the coefficient of $\psi_A^2$ in the energy becomes negative. This occurs once the inverted state enters the \textbf{amplification region} (discussed in the main text) given by
\begin{equation}
\label{eqn:coeff}
   f(\mu,\psi_S)<0\,\,\Rightarrow\,\,\,\psi_S^2<\frac{b_2\mu-b_1}{b_3},
\end{equation}
and thus becomes unstable to asymmetric modes. At this point, the system snaps through via asymmetric modes to the only stable equilibrium (minimum): the natural state. 
Therefore, a quasi-static analysis suggests that our double-mass von Mises truss becomes unstable to asymmetric modes first, like an Elastica. 

If we consider dynamic effects, however, the behavior changes. If the end distance is changed fast, then the (dynamic) value of $\psi_S$ lags behind the actual value of the equilibrium $\psi_I$ due to inertial effects. This means that $\psi_S^2>\psi_I^2$, resulting in a larger value of the coefficient $f(\psi;\mu)$ shown in eqn \eqref{functionf}. If the dynamics is sufficiently fast, the system lags behind enough to never enter the amplification region and thus never becomes unstable to asymmetric modes. However, a (delayed) snap-through still occurs when $\mu<\mubif$, as the inverted state ceases to exist altogether in a saddle-node bifurcation and the system is forced to jump to the natural state. 

Therefore, dynamic effects determine whether the system enters or not the amplification region and thus characterise both the symmetry and time of the snap-through. To make this claim more precise, we need to derive the equations of motion and solve for the path of the system.

\begin{figure}[t]
\includegraphics[width=\columnwidth]{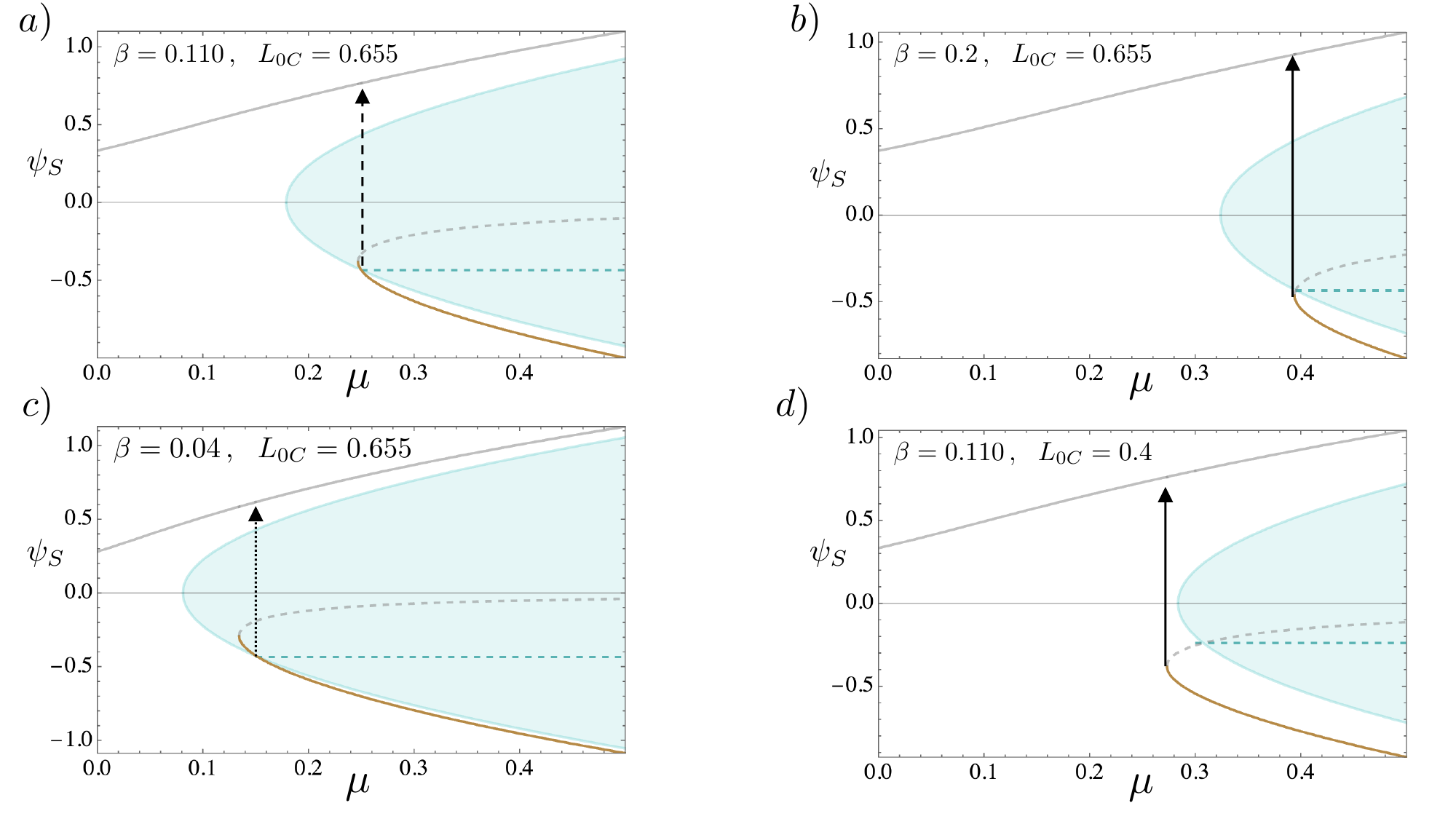}
\caption{The bifurcation diagram of the double-mass von Mises truss for different values of the parameters $\beta$ and $L_{0C}$. a) The parameters are chosen so that the pitchfork bifurcation in $\psi_A$ and the saddle node bifurcation in $\psi_S$ occurs exactly at the same value of $\mu$ as in the continuous arch problem. Panels b), c) and d) show the bifurcation diagrams for systems with larger bending stiffness, smaller bending stiffness and smaller distance between the central masses respectively. Vertical lines show the  path for quasi-static snap-through, dashed curves show branches in which the system is unstable to an asymmetric perturbation, while solid curves denote stable regions, for which stability is only lost due to a saddle node (symmetric) bifurcation. In each panel, the cyan region is the amplification region, in which $f<0$.}
\label{fig:parametersDiagram}
\end{figure}

The nature and stability of the system depends on our choice of the two parameters $\beta$ and $L_{0C}$.  To run all our numerical calculations, and to discuss the general stability of our system, we have chosen $\beta$ and $L_{0C}$ so that the bifurcation diagram of the double-mass von Mises truss is as close as possible to that of the shallow arch derived in \S\ref{sec:SI:Bifurcation}. In particular, we require that the saddle-node bifurcation point described in equation \eqref{eqn:saddlenode} occurs at $\mu_*=0.247$ (i.e.~we ensure that $\mubif^{\mathrm{vM}}=\mubif$) and that the loss of stability to asymmetric modes found in equation $\eqref{eqn:pitchfork}$ occurs at $\mu_A^{\mathrm{vM}}=\mu_A=1/4$. These two conditions lead to a choice of $\beta=0.110$  and $L_{0C}=0.655$.

It is important to note that a different choice of parameters may lead to a different bifurcation diagram and, possibly, to a different behaviour.
In figure \ref{fig:parametersDiagram}, we show the bifurcation diagram that replicates the arch problem (a) as discussed in the previous paragraph, and three other diagrams with a different choice of parameters. Recall that, when reading these diagrams, one starts from the right and reduces the end-distance $\mu$ by moving towards the left of the plot. Panel b) shows a system with large bending stiffness. Clearly, the whole bifurcation diagram is shifted to the right. Furthermore, unlike in a), the system looses stability due to the saddle node bifurcation, with the asymmetric pitchfork occurring only on the unstable branch (dashed gray) not explored by the system. Conversely, in c), we show what happens for small bending stiffness, with the whole diagram shifted to the left and the system becoming unstable to asymmetric perturbations first before the saddle-node bifurcation occurs. Importantly, and perhaps counter-intuitively, even if a system goes unstable due to a saddle node bifurcation (and thus initially in via a symmetric mode), it does not mean the transient path to the new stable state will remain symmetric. Indeed, as discussed in the main text and in detail in the next section of this SI, the appearance of asymmetry during the transient jump depends on whether or not the system traverses the amplification region. Panels a), b) and c) show system in which th jump always traverses this region (cyan). Conversely, in panel c) we show an example of a system that jumps and never enters the amplification region. Such a system will not see amplification of the asymmetric mode.  

In the simple toy model, the condition for our trajectories to never traverse the amplification region is that, in our bifurcation diagrams, the saddle-node point is on the left of (any point in) the amplification region. The leftmost point of the asymmetric region occurs at $\psi_S=0$ which, when substituting into $f$ and setting everything equal to zero tells us that it coincides with $\mu=\frac{2 \left(2 \beta +\beta  L_{0C}^2+2 \beta  L_{0C}\right)}{L_0 \left(L_{0C}+2\right){}^2}$. Equating value of $\mu$ to the position of the saddle node given in eqn. \eqref{eqn:saddlenode} tell us that asymmetric amplification is observed provided 
\begin{align}
    \beta < \frac{27}{256} L_{0C}^3 (L_{0C}+2)^3.
\end{align}

\section{Dynamics of the double-mass von Mises truss}
In this section, we supplement the elastic energy already considered with the kinetic energy of the system. This allows us to study the dynamics and thereby elucidate how the rate $|\dot{\mu}|$ at which the end-to-end displacement is decreased modifies the behavior of the system. Here, we show that for sufficiently small $|\dot{\mu}|$ (almost quasi-static variations), the double-mass von Mises truss  snaps from the inverted to the natural state via an asymmetric mode. However, for larger $|\dot{\mu}|$, the importance of asymmetric modes to promote the snap through decreases until, for $|\dot{\mu}|>|\dot{\mu}|_c$, the double-mass von Mises truss always goes unstable symmetrically.

\subsection{Deriving the equations of motion\label{sec:TimeScaling}}

Since we have already derived the potential energy of the system, it is convenient to write down the kinetic energy and then use Lagrangian mechanics to derive the equations of motion. The kinetic energy is easily written down by considering the motion (in polar coordinates) of the masses about the two edge elements:
\begin{equation}
    T=\half m \left[(l_L \dot{\theta}_L)^2+\dot{l}_L^2+(l_R \dot{\theta}_R)^2+\dot{l}_R^2\right]
\end{equation}
where dots indicate derivatives with respect to time $t$. We can now use the rescalings from equation \eqref{eqn:rescalings}. 
We also want to choose an appropriate rescaling of time. One natural choice to do so and try and match the results from the simple model with the full problem, is to rescale time by the typical time scale of oscillations of the system. We shall calculate this time scale later. For now, we simply rescale time as:
\begin{equation}
    t=\tosc \left( \sqrt{\frac{m}{k}}\frac{1}{\alpha}\right)\tau. 
\end{equation}
where $ \sqrt{m/k}/ \alpha$
is the time scale of the oscillations of the linear springs, $\tau$ is the dimensionless time and $\tosc$ is a factor which will be determined later by studying the period of oscillations in the natural state.

Expanding the kinetic energy in small $\alpha$ we can write the rescaled kinetic energy $\tilde{T}=\frac{T}{k l_0^2 \alpha^4}$ as
\begin{equation}
\tilde{T}=\frac{1}{{\tosc}^2}\left(\dot{\psi}_S(\tau)^2+\fourth\dot{\psi}_A^2+H.O.T. \right)
\label{eqn:ke}
\end{equation}
where now, in a slight abuse of notation, dots refer to derivative with respect to the rescaled time $\tau$. Note that the kinetic energy scales like the potential energy in $\alpha$ and the two are thus comparable. The terms in the kinetic energy can be explained by realizing that, for small angles, only the vertical velocities of the masses matter. This is because the change in vertical displacements scale like $\alpha^2$ (from the scaling in equation \eqref{eqn:rescalings}) while vertical displacements scale like $\alpha$ giving $y_L=l_0 \sin(\alpha (\psi_S + \half \psi_A)) \sim l_0\alpha (\psi_S + \half \psi_A)$ and $y_R=l_0\sin(\alpha (\psi_S -\half  \psi_A)) \sim l_0\alpha (\psi_S - \half \psi_A)$. Differentiating these positions with respect to time leads to $\dot{y_L} \sim \alpha^2 \sqrt{k/m} l_0(\dot{\psi_S} +\half \dot{\psi_A})/\tosc $ and $\dot{y_R} \sim \alpha^2 \sqrt{k/m} l_0(\dot{\psi_S} - \half \dot{\psi_A})/\tosc$. Substituting in the kinetic energy $\half m (\dot{y}_L^2+\dot{y}_R^2)$ and rescaling yields the result in equation \eqref{eqn:ke}.

We write the (rescaled) Lagrangian of the system $L=\tilde{T}-\tilde{U}$, where $\tilde{U}$ is given in eqn.~\eqref{eqn:energy2}. We minimize the action with respect to variations in $W_S$, $W_A$, $\psi_S$ and $\psi_A$. Minimization with respect to $W_S$ and $W_A$  leads to \eqref{eqn:WS} and \eqref{eqn:WA}, and thus yields no new information.  Minimization with respect to $\psi_A$ and $\psi_S$ leads to the 
\textbf{equations of motion}:
\begin{align}
\label{eqn:dyn1}
    \ddot{\psi}_S/{\tosc}^2&= \half a_1 - \half(a_2-a_3 \mu)\psi_S-  \half a_4 \psi_S^3-\half b_3 \psi_S \psi_A^2\\
        \label{eqn:dyn2}
    \ddot{\psi}_A/{\tosc}^2&=-2 f(\psi_S,\mu) \psi_A-2 b_4 \psi_A^3
\end{align}
where the right-hand side of both equations is just the derivative of $\tilde{U}$ in eqn.~\eqref{eqn:simpleenergy} with respect to $\psi_S$ and $\psi_A$ for the first and second equations, respectively. For small $\psi_A$, these equations coincide with eqns.~(5) and (6) of the main text.

Unfortunately, these equations are non-linear and cannot be solved explicitly. Furthermore, the system does not conserve energy: substituting an explicit time dependence of the end-shortening  on time, shows that the Lagrangian $L$ depend explicitly on time.
Note that in the main text, we substitute $\mu=\mu_A-|\dot{\mu}|\tau$ so that the quasi-static instability occurs at $\tau=0$. For the numerical calculations, we change this convention and we write $\mu=\mu_i-|\dot{\mu}|\tau$ where $\mu_i$ is our initial end-shortening. This choice is made for convenience and should not cause concern since all numerical results are plotted with respect to $\mu$, not $\tau$, making any shift in time irrelevant.

\subsubsection{Oscillations about the natural state}

Finally, we can find the typical period of oscillations about the natural state for the toy model. To do so, we write:

\begin{align}
    \psi_S(\tau)&=\psi_N+\varepsilon \cos(\omega_S \tau)\\
    \psi_A(\tau)&=\varepsilon \cos(\omega_A \tau)
\end{align}
where $\psi_M$ is the natural state equilibrium found in \eqref{eqn:EquilibriaSimple}. If we linearize the equations of motion \eqref{eqn:dyn1} and \eqref{eqn:dyn2} we get the two equations for $\omega_S$ and $\omega_A$:
\begin{align}
  (\omega_S /\tosc)^2&=  a_2+3 a_4 \psi_0^2-a_3 \mu\\
  (\omega_A /\tosc)^2&=  b_1+3 b_3 \psi_0^2-b_2 \mu.
\end{align}
Like for the arch case, the natural choice is to evaluate the ringing frequency at the same end-shortening at which the system goes first unstable: $\mu=1/4$. Substituting this value in and our values for the coefficients, we find that $(\omega_A /\tosc)=0.663$ and $(\omega_S /\tosc)=0.734$. Since we are looking for the lowest frequency oscillations, we take this as the frequency of oscillations. Finally, we find that the time period of this oscillation is 
\begin{equation}
\tosc=2 \pi/\omega_s=9.477.
\end{equation}

\subsubsection{Initial conditions}
Now that we have fixed a time scale, we can plot the trajectories of the system. This can only be done numerically. However, to do so, some care has to be taken when imposing the initial conditions. 

In principle, we could choose any reasonable initial condition, as for instance $\psi_S(\tau_i)=\psi_i$, $\dot{\psi}_S(\tau_i)=0$ and $\psi_A(\tau_i)=\psi^0_A$, $\dot{\psi}_A(\tau_i)=0$. However, our choice of initial condition will affect the trajectory of the system, thus adding extra variables to the problem. For example, choosing an initial $\psi_S$ that is far form the equlibrium inverted solution, would introduce large oscillations that may heavily affect the dynamics. Similarly, starting on the equilibrium branch, but with a `wrong' initial $\dot{\psi}_S$, may also introduce large oscillations and thus the same issue.

To get around this problem, we require the solution to the equations of motion to map back to the stable inverted solution as $\tau\to-\infty$, i.e.~at a large time prior to the instability. (This is analogous to the method used to find the correct initial conditions when the control parameter $\mu$ varies dynamically to control snap-through \cite{Liu2021}). Imposing such a condition on both equations of motion is difficult. We can assume that the amplitude of the asymmetric component of motion is much smaller than that of the symmetric one, and ignore the contribution of $\psi_A$ for large negative times. We then need to find an expansion for $\tau \to -\infty$ of 
\begin{equation}
    \ddot{\psi}_S/{\tosc}^2=\half a_1-\half(a_2-a_3 \mu) \psi_S-\half a_4 \psi_S^3.
\end{equation}
We expand $\psi_S$ in powers of $\sqrt{-\tau}$ and solve order by order. Substituting in the definitions of the constants $a_1$, $a_2$, etc.~we find that:
\begin{align}
\psi_S=-\sqrt{(2+L_{0C})}\sqrt{|\dot{\mu}|}\sqrt{-\tau}+\frac{\beta}{\sqrt{(2+L_{0C})}\sqrt{|\dot{\mu}|}\sqrt{-\tau}}-\frac{\beta}{2(2+ L_{0C})|\dot{\mu}| \tau}+... .
\end{align}
Correspondingly, the first derivative of $\psi_S$ for large negative times is:
\begin{align}
\dot{\psi}_S=\frac{\sqrt{(2+L_{0C})}\sqrt{|\dot{\mu}|}}{2 \sqrt{-\tau}}-\frac{\beta}{2\sqrt{(2+L_{0C})}\sqrt{|\dot{\mu}|}(-\tau)^{3/2}}+\frac{\beta}{2(2+ L_{0C})|\dot{\mu}| \tau^2}+... .
\end{align}

In the numerical solutions presented in figure 4 of the main text, we use the first two terms in the above series to determine $\psi_S(\tau_i)$ and $\dot{\psi}_S(\tau_i)$ evaluated at initial time $\tau_i=-100$ to impose boundary conditions on $\psi_S$. This choice ensures that the initial error on $\psi_S$ is of order $10^{-4}$ while the error on its derivative is of order $10^{-6}$, as evaluated from the first discarded term in the above series. We find this approximation sufficient so that the numerical solution correctly traces back to the equilibrium branch at large negative times without any unwanted oscillations.

The boundary conditions for $\psi_A$ are considerably simpler: we just impose $\dot{\psi}_A(\tau_i)=0$ and an initial perturbation $\psi_A(\tau_i)=\psi^0_A$.

\subsection{Amplification and phase of the oscillation}
\begin{figure}[t]
\includegraphics[width=0.8 \columnwidth]{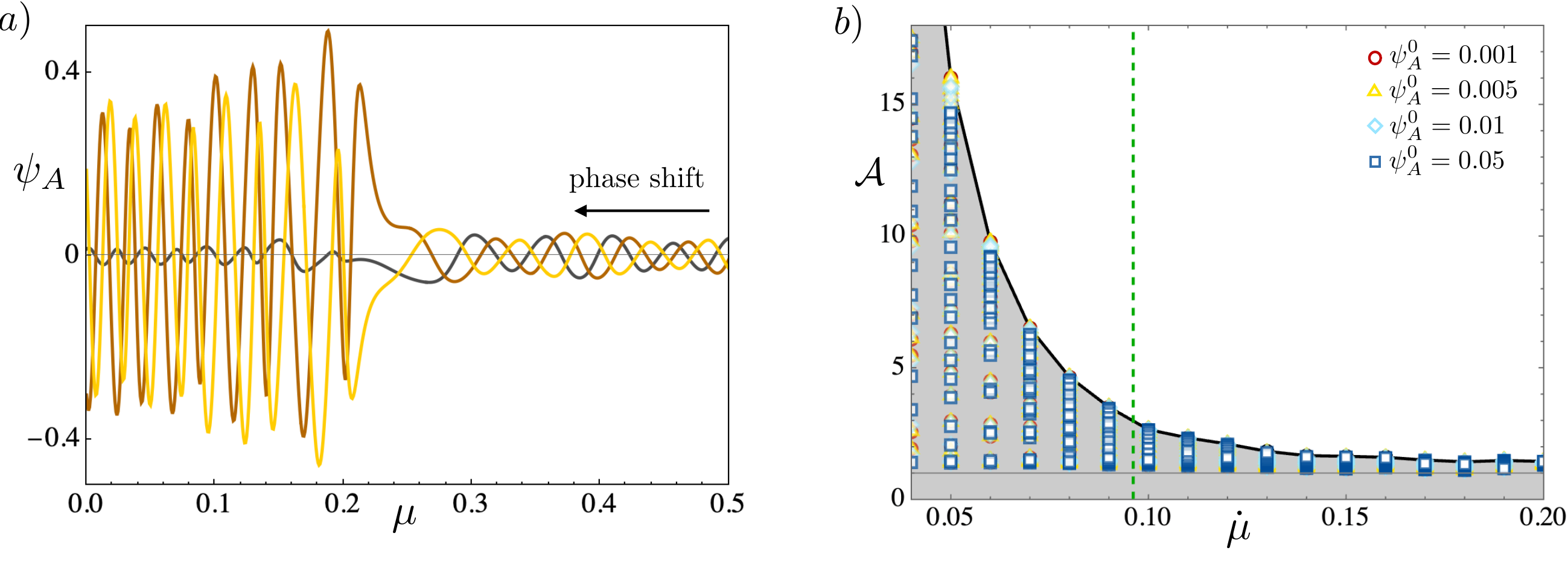}
\caption{a) The asymmetric angle $\psi_A$ for simulations with $\dot{\mu}=0.03$ but different phase shifts. Note that the black path has minimal amplification simply because of the phase. b) Plot of all amplifications obtained at different phases and different rates $\dot{\mu}$. The envelope (gray region) is found by taking the maximum amplification  of all simulations performed at a given rate. } 
\label{fig:amplification}
\end{figure}

For each end-shortening rate $\dot{\mu}$, we may measure the maximum angle $\psi_A$. In doing so, we note that the maximum angle depends heavily on the phase of the oscillation as the system traverses the amplification region. As shown in figure \ref{fig:amplification} even at the same speed, our path will look very different depending on the phase. Crucially, certain paths will enter the amplification region with a certain trajectory which means that little to no amplification of the oscillation occurs. 

To capture the amplification potential (the maximum possible amplification) with a given loading rate $\dot{\mu}$ in the toy model, we compute 25 different trajectories, each with a different phase (spanning roughly one period of oscillation). Some typical examples of the resulting trajectories are shown in figure \ref{fig:amplification}(a), while a summary of the resulting variety of amplifications observed is shown in figure \ref{fig:amplification}(b). To produce figure 4 in the main text, we have chosen the maximum amplification observed at each $\dot{\mu}$. 

\subsection{Videos of the motion}
The supplementary information includes two videos of the motion with different values of $\dot{\mu}$. In the first video, $\dot{\mu}=0.01<\dot{\mu}_c$, with the resulting video accelerated ten-fold for clarity; in this case, $\mu$ changes  relatively slowly and so a small initial asymmetry is amplified. In the second video, $\dot{\mu}=\dot{\mu}_c$ and  the system skirts the amplification region without the initial asymmetry being amplified significantly. 

\section{Critical path}
As discussed in the main text, the initial asymmetry in the system is  amplified only when the trajectory of the system enters the \textbf{amplification region}, which happens when $|\dot{\mu}|>|\dot{\mu}_c|$. In this section, we determine the critical rate $|\dot{\mu}_c|$ that marks this transition between trajectories that do and do not enter the amplification region. 

\subsection{Approximation near the saddle-node bifurcation point}

When the perturbation in the asymmetric mode is small, the dynamic system can be simplified to:
\begin{align}
\label{eqn:dyn1B}
    \ddot{\psi}_S/{\tosc}^2&=\half a_1 - \half(a_2-a_3 \mu)\psi_S- \half a_4 \psi_S^3\\
        \label{eqn:dyn2B}
    \ddot{\psi}_A/{\tosc}^2&=-2(b_1 -b_2 \mu+b_3 \psi_S^2) \psi_A
\end{align}

It is instructive to first write eqn.~\eqref{eqn:dyn1B} in terms of the underlying parameters, $\mu$ and $\beta$; this leads to:
\begin{equation}
3 \ddot{\psi}_S=\beta - \left((2 + L_{0C})\mu-2\beta \right)\psi_S-  \psi_S^3 .
\label{eqn:dyn1C}
\end{equation}
We can approximate the trajectory near the saddle-node bifurcation point given by $\psi_S^*$ and $\mu_*$ defined in equation \eqref{eqn:saddlenode}. To do so, we let:
\begin{align}
\psi(\mu)&=\psi^*_S+\left(\frac{|\dot{\mu}|}{\tosc}\right)^{2/5}\left(\frac{2 }{\beta}\right)^{1/15} \left(\frac{2-L_{0C}}{3}\right)^{2/15}  \delta \psi_S(\Delta)\\
\mu&=\mubif+\left(\frac{|\dot{\mu}|}{\tosc}\right)^{4/5}\left(\frac{2 }{\beta}\right)^{2/15}\left(\frac{3}{2-L_{0C}}\right)^{1/15}\Delta
\label{eqn:rescalingsd}
\end{align} 
and keep only the leading order terms in $\delta \psi_S$ and $\Delta$ so that eqn. \eqref{eqn:dyn1B} takes the simple form:
\begin{align}
\label{eqn:dyn1C}
    \frac{d^2 }{d \Delta^2}\delta \psi_S&=\delta \psi_S^2(\Delta)-\Delta.
\end{align} 
This equation can easily be solved numerically and describes the trajectory of our system near the saddle-node bifurcation for all parameters $\beta$, $L_{0C}$ and $|\dot{\mu}|$.

Next, we focus on understanding when the trajectory of the system enters the amplification region. To do so, we look at the function $f$ in eqn.~\eqref{functionf}. We are interested in understanding which values of $|\dot{\mu}|$ result in $f$ changing sign and becoming negative, thus amplifying the asymmetric perturbation. We substitute the new variables from eqn.  \eqref{eqn:rescalingsd} in the definition of $f$ and obtain
\begin{align}
f(\Delta)=- c_0-c_1 \left[3\left(\frac{|\dot{\mu}|}{\tosc}\right) ^{4/5} \Delta - \left(\left(\frac{|\dot{\mu}|}{\tosc}\right)^{2/5} \delta \psi_S -c_2\right)^2\right],
\end{align}
where 
\begin{align}
    c_0=\frac{3 L_{0C} (2 \beta^2)^{1/3}(2+L_{0C})-8 \beta}{12 L_{0C}^2}, \quad \quad \\
    c_1=\left(\frac{(2+L_{0C})}{3}\right)^{9/5}\frac{1}{2^{13/15} L_{0C} \beta^{2/15}}, \quad c_2=\left(\frac{3\beta}{2(2+L_{0C})}\right)^{2/5}.
\end{align}

The critical path, where the trajectory only touches but never enters the amplification region, can be found by solving $f(\Delta)=0$ and $f'(\Delta)=0$ simultaneously. We can easily do this numerically and we obtain that the critical rate is  $|\dot{\mu}_c| =0.096$, as discussed in the main text. When $|\dot{\mu}|<|\dot{\mu}_c|$, the trajectory enters the bifurcation region and amplification of the asymmetric mode is observed.


\providecommand{\noopsort}[1]{}\providecommand{\singleletter}[1]{#1}%